\documentclass[11pt]{article}
\pdfoutput=1
\usepackage[margin=1.0in]{geometry}
\usepackage{amsmath,amssymb,graphicx}
\usepackage{bm}
\usepackage{hyperref}
\usepackage{slashed}
\usepackage[utf8]{inputenc}

\usepackage[T1]{fontenc}
\usepackage{baskervald}

\usepackage[font=small,labelfont=bf]{caption}
\usepackage{cite}

\newcommand{\be}{\begin{equation}}
\newcommand{\ee}{\end{equation}}

\DeclareMathOperator\sign{sign}

\newcommand{\tB}{{\widetilde B}}
\newcommand{\tW}{{\widetilde W}}
\newcommand{\tH}{{\widetilde H}}
\newcommand{\tg}{{\widetilde g}}
\newcommand{\tchi}{{\widetilde \chi}}

\newcommand{\tildet}{{\widetilde t}} 
\newcommand{\sigmabar}{{\overline \sigma}}

\newcommand{\GeV}{\,\mathrm{GeV}}
\newcommand{\TeV}{\,\mathrm{TeV}}

\newcommand{\iu}{{\mathrm i}}

\usepackage[usenames,dvipsnames]{xcolor}
\definecolor{darkerblue}{rgb}{0.2,0.2,0.5}

\usepackage{afterpage}

\usepackage{tikz}
\usetikzlibrary{trees}
\usetikzlibrary{decorations.pathmorphing}
\usetikzlibrary{decorations.markings}
\tikzset{
    photon/.style={decorate, decoration={snake}, draw=black},
    wino/.style={draw=redwine},    
    electron/.style={draw=black, postaction={decorate},
        decoration={markings,mark=at position .55 with {\arrow[draw=black]{>}}}},
    scalar/.style={draw=black, dashed,postaction={decorate},
        decoration={markings,mark=at position .55 with {\arrow[draw=black]{>}}}},
    gluon/.style={decorate, draw=black,
        decoration={coil,amplitude=4pt, segment length=5pt}}
}
\tikzstyle{blob}=[circle,
                              thick,
                              minimum size=0.4cm,
                              draw=gray!80,
                              fill=gray!60]   
\tikzstyle{redblob}=[circle,
                              thick,
                              minimum size=0.4cm,
                              draw=red!80,
                              fill=red!60] 
                              
\def\Smiley{\tikz[baseline=-0.75ex,black]{
    \draw circle (2mm);
    \node[fill,circle,inner sep=0.5pt] (left eye) at (135:0.8mm) {};
    \node[fill,circle,inner sep=0.5pt] (right eye) at (45:0.8mm) {};
    \draw (-155:1.25mm) arc (-130:-50:1.8mm);
    }
}

\def\Sadey{\tikz[baseline=-0.75ex,black]{
    \draw circle (2mm);
    \node[fill,circle,inner sep=0.5pt] (left eye) at (135:0.8mm) {};
    \node[fill,circle,inner sep=0.5pt] (right eye) at (45:0.8mm) {};
    \draw (-155:1.25mm) arc (130:50:1.8mm);
    }
}                      

\title{\bf Deciphering the MSSM Higgs Mass at Future Hadron Colliders }

\author{Prateek Agrawal$^a$, JiJi Fan$^b$, Matthew Reece$^{a,c}$, and Wei Xue$^d$\\
{\small $^a$ \em Department of Physics, Harvard University, Cambridge, MA 02138, USA}\\
{\small $^b$ \em Department of Physics, Brown University, Providence, RI 02912, USA}\\
{\small $^c$ \em School of Natural Sciences, Institute for Advanced Study, Princeton, NJ 08540, USA}\\
{\small $^d$ \em Center for Theoretical Physics, Massachusetts Institute of Technology, Cambridge, MA 02139, USA}}

\begin{document}
\maketitle

\begin{abstract}
Future hadron colliders will have a remarkable capacity to discover massive new particles, but their capabilities for precision measurements of couplings that can reveal underlying mechanisms have received less study. In this work we study the capability of future hadron colliders to shed light on a precise, focused question: is the higgs mass of 125 GeV explained by the MSSM? If supersymmetry is realized near the TeV scale, a future hadron collider could produce huge numbers of gluinos and electroweakinos. We explore whether precision measurements of their properties could allow inference of the scalar masses and $\tan \beta$ with sufficient accuracy to test whether physics beyond the MSSM is needed to explain the higgs mass. We also discuss dark matter direct detection and precision higgs physics as complementary probes of $\tan \beta$. For concreteness, we focus on the mini-split regime of MSSM parameter space at a 100 TeV $pp$ collider, with scalar masses ranging from 10s to about 1000 TeV.
\end{abstract}

\section{Introduction}
\label{sec:intro}

Even as the LHC probes the TeV energy scale, a significant effort is underway to plan for future hadron colliders at higher energies \cite{Avetisyan:2013onh, Richter:2014pga, Rizzo:2015yha, Benedikt:2015poa, Tang:2015qga, Arkani-Hamed:2015vfh, CEPC-SPPCStudyGroup:2015csa, Mangano:2016jyj, Golling:2016gvc, Contino:2016spe}. The Large Hadron Collider has given us two major clues so far about the nature of physics at higher energies: the discovery of the higgs boson with mass 125 GeV and the absence of any significant evidence for new particles. These results have forced the high-energy theory community to reevaluate the most compelling models explaining the origin of the electroweak scale, such as weak-scale supersymmetry. Nevertheless, SUSY persists. The data seems to point to simpler models where the weak scale is ``meso-tuned'' rather than more elaborate natural models which obtain the correct higgs mass through an extended mechanism. Future colliders at higher energies hold a lot of promise to probe these well-motivated models.

The earliest studies of supersymmetry at future colliders have focused on the mass scales that can be probed at 33 and 100 TeV proton--proton colliders. Broadly speaking, a 100 TeV collider can discover colored particles with masses near 10 TeV \cite{Borschensky:2014cia, Stolarski:2013msa, Cohen:2013xda, Jung:2013zya, Cohen:2014hxa, Ellis:2015xba} and electroweak particles with masses near 1 TeV \cite{Low:2014cba, Gori:2014oua, Bramante:2014tba, Bramante:2015una, Ismail:2016zby}. Of course, it is not surprising that the mass reach of a collider operating at 10 times the LHC energy can probe particles an order of magnitude heavier than those the LHC probes. Quantifying this reach is a crucial first step; here we emphasize a complementary point of view. It is important to formulate precise physical questions that can lend insights into new mechanisms, and evaluate  the capability of future colliders to answer them. Such studies will provide valuable input to the design of these colliders. Therefore, even at this early design phase it is vital to look beyond the discovery reach and study the potential of future colliders to address fundamental questions. 
For future hadron colliders, a partial list of important qualitative questions that have been considered to date include whether dark matter arises from an SU(2) multiplet \cite{Low:2014cba, Bramante:2014tba, Bramante:2015una}, how the Higgs boson interacts with itself \cite{Azatov:2015oxa,Yao:2013ika,Barr:2014sga,He:2015spf,Contino:2016spe}, whether the electroweak phase transition was first order \cite{Curtin:2014jma,Assamagan:2016azc}, and how the Standard Model behaves in the electroweak-symmetric regime \cite{Hook:2014rka}. 

Our goal in this paper is to add a new qualitative question to the list of physics goals for a future hadron collider: does the MSSM explain the observed higgs boson mass of 125 GeV? The MSSM has the virtue that the higgs mass is calculable: it is predicted in terms of measurable supersymmetry-breaking effects. If evidence hinting at a supersymmetric spectrum emerges---for instance, if a color-octet fermion that could be a gluino is discovered---then, in order to assess whether the MSSM is actually responsible for the underlying physics, we must measure the properties of the newly discovered particles more extensively. If the gluino mass is in the TeV range, a future hadron collider will be a gluino factory. For example, at a 100 TeV collider, 3 ab$^{-1}$ of data would contain about 20 million gluino pair events if the gluino mass is 2 TeV and a hundred thousand events if the mass is 5 TeV \cite{Borschensky:2014cia}. Such large event rates will allow the accurate measurement of gluino branching ratios, even of rare decays. The gluinos will cascade through various electroweakinos, which are also produced directly. The goal of our work is to develop observables that allow us to measure the properties of these fermionic particles accurately enough to test the MSSM higgs mass prediction. 

\begin{figure}[!h]\begin{center}
\includegraphics[width=0.45\textwidth]{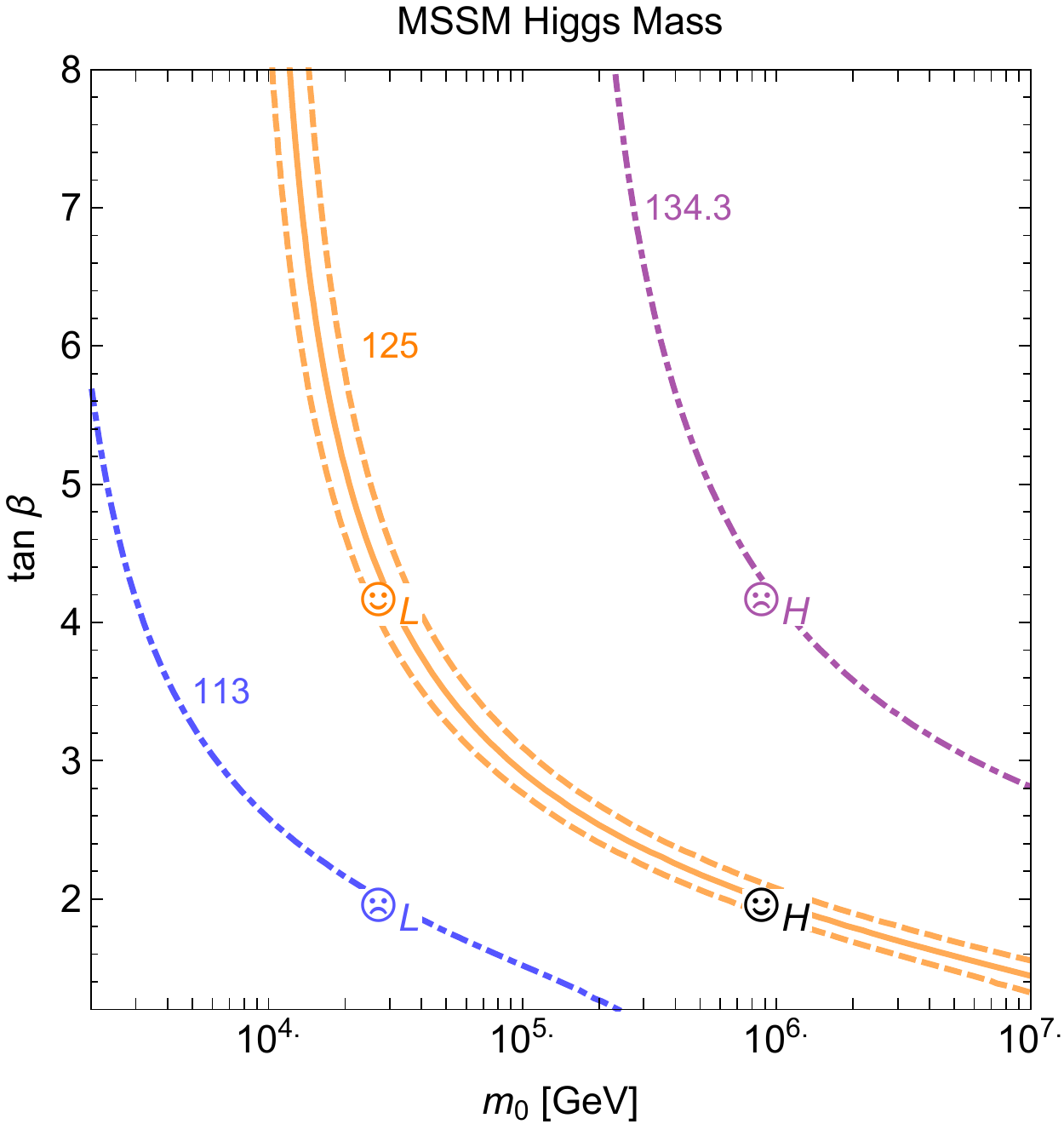}
\end{center}
\caption{Contours of the MSSM higgs boson mass, predicted in terms of a universal scalar mass $m_0$ and the higgs VEV ratio $\tan \beta$. We have computed the mass using SusyHD \cite{Vega:2015fna} with the choice $A_t = 0, m_{1/2} = 1$ TeV. (The answer is not very sensitive to the fermion masses.) The observed mass of 125 GeV is indicated by the solid orange curve, bracketed by dashed orange curves indicating theoretical uncertainty. Parameter ranges giving answers differing in either direction by about 10 GeV are indicated by the dot-dashed purple and blue curves. Four points are singled out for further study in examples: two \protect\Smiley points with the correct higgs mass and two \protect\Sadey points with the wrong higgs mass.}
\label{fig:higgsmass}
\end{figure}%

As is well-known, at tree level the MSSM predicts $m_h < m_Z$, but  loop corrections can raise the higgs mass \cite{Barbieri:1990ja,Haber:1990aw,Casas:1994us,Carena:1995bx}. A great deal of effort has gone into multi-loop computations of the higgs mass in the MSSM, as reviewed in \cite{Draper:2016pys}. We can expect that by the time a future hadron collider is operational, the theoretical uncertainties will be further reduced. Although a high-precision check of the MSSM may require a more detailed solution of the SUSY inverse problem, measurement of the stop masses $m_{\tilde t}$, the stop mixing $A_t$, and the higgs VEV ratio $\tan \beta$ allows an approximate check. These determine the dominant one-loop threshold corrections to the higgs boson quartic coupling and hence the mass of the physical higgs. In the case $A_t = 0$, the dependence of the higgs mass on the other parameters is shown in Figure \ref{fig:higgsmass}. If the MSSM is correct, we expect measurements to land near the orange curves, while a measurement elsewhere in the plane would indicate either physics beyond the MSSM or a substantial role for the parameter $A_t$. To illustrate some possibilities, we have indicated two points marked with the symbol {\large \Smiley}, one at $m_0 \approx 30$ TeV and $\tan \beta \approx 4$ (labeled $L$ for ``low mass,'' comparatively speaking!) and one at $m_0 \approx 1000$ TeV and $\tan \beta \approx 2$ (labeled $H$ for ``high mass''). If we exchange the $(m_0, \tan \beta)$ pairings, we obtain two other points marked with {\large \Sadey}, for which the MSSM predicts a higgs mass that is wrong by more than 10 GeV. As a crude test of whether a future collider can test the MSSM, we can ask whether it could distinguish among these four points at high significance.

Both of the {\large \Smiley} points we have chosen lie in the
``meso-tuned'' regime; they do not fully solve the hierarchy problem,
though supersymmetry would still explain most of the hierarchy,
leaving a residual fine-tuning unexplained. In this regime,
the MSSM may be the correct theory even though the mass scales we would like to probe are likely
to be out of reach of even the next generation of high-energy
colliders. First-generation squarks with masses near 30 TeV may be
probed in associated production with a gluino \cite{Ellis:2015xba},
though stops near the same mass would be out of reach. Our challenge
will be to test the scalar mass scale indirectly, given the gluinos
and electroweakinos that we expect to have access to if the SUSY spectrum
is somewhat split. Another region of MSSM parameter space has lighter
stops, perhaps even near the TeV scale, with large $A_t$. In this
region, we could hope to measure the stop masses and $A_t$ directly
(for instance, along the lines discussed in \cite{Perelstein:2007nx,
Blanke:2010cm}). We will not linger on the case of light stops and
large $A_t$ in this paper, focusing instead on the case of a
moderately split spectrum where we have access only to fermionic
superpartners.

The benchmark values of scalar masses at 30 TeV and 1000 TeV are well motivated from a theoretical point of view. A 30 TeV mass scale for particles that interact with gravitational strength, like gravitinos and moduli, allows them to decay just before BBN \cite{Coughlan:1983ci,Ellis:1986zt,deCarlos:1993wie,Moroi:1995fs,Kane:2015jia}, ameliorating cosmological problems. In many such models, the masses of squarks and sleptons will be at the same scale as the gravitino mass while the gauginos are lighter by roughly a loop factor. This is true both in anomaly mediation with unsequestered scalars \cite{Randall:1998uk,Giudice:1998xp,Wells:2003tf} and in some incarnations of moduli mediation \cite{Choi:2005ge,Choi:2005hd,Acharya:2007rc,Acharya:2008zi}. The case with scalars at 1000 TeV is also well-motivated. If we study split SUSY scenarios where the scalar masses at the GUT scale are universal, we will find small $\tan \beta$ because $m^2_{H_u} = m^2_{H_d}$ in the ultraviolet. RG running in this case pushes $\tan \beta$ up to about 2, calling for 1000 TeV scalars in order to achieve $m_h \approx 125$ GeV \cite{Bagnaschi:2014rsa}. Furthermore, the 1000 TeV scale emerges in certain large-volume sequestering scenarios \cite{Blumenhagen:2009gk,Aparicio:2014wxa,Reece:2015qbf} with approximate no-scale structure \cite{Cremmer:1983bf,Ellis:1983sf,Balasubramanian:2005zx,Conlon:2005ki}. Hence, a variety of top-down considerations point to both the benchmark points, with scalars near 30 and 1000 TeV, and it would be of great interest to determine if either case is realized in nature. This is a strong motivation for attempting to measure the scalar mass scale even when the scalars themselves are beyond the direct reach of our colliders. 

Having motivated the problem of measuring the scalar mass scale $m_0$
and $\tan \beta$ from observations purely involving fermionic
superpartners, we will turn our attention to the experimental
observables that are indirectly sensitive to these parameters. In
\S\ref{sec:gluinoobservables} we will discuss observables associated
with gluinos. In particular, a one-loop gluino decay is sensitive to
$m_0$; decays to higgsinos are sensitive to $\tan \beta$. In
\S\ref{sec:ewkobservables} we will discuss how to use observables
associated with electroweak states to measure $\tan \beta$. In this
case there are a number of probes, including electroweakino decay
branching ratios, higgs boson decays, and dark matter direct
detection. We discuss the prospects for such measurements, and outline
which are likely to be most effective depending on the ordering of
bino, wino, and higgsino masses in the spectrum. In
\S\ref{sec:colliderstudy}, we present an example case study for how to
measure both $m_0$ and $\tan \beta$ at a 100 TeV collider for a
spectrum with the mass ordering $M_3 > M_2 > \mu > M_1$. In
\S\ref{sec:conclusions} we offer some concluding remarks.

Note: we have previously contributed an early version of this work as
\S3.10 of the 100 TeV BSM study \cite{Golling:2016gvc}. The collider
case studies presented there are different from those presented here;
they cover different electroweakino spectra, and did not include a
study of Standard Model backgrounds. Here we present one example
collider case study, different from those in the earlier study, for
which we can include SM backgrounds and present a final estimate of
error bars for the measurement in the $(m_0, \tan \beta)$ plane.

\section{Observables from gluino decays}
\label{sec:gluinoobservables}

Given their large production cross sections, gluinos are promising
candidates to measure the scalar mass scale and $\tan\beta$. At low scalar masses, we could attempt to measure the scalar mass directly through pair production, but this approach would not work beyond 10 TeV in stop mass \cite{Cohen:2014hxa}. First-generation squarks can be produced from a valence quark in the process $q g \to \widetilde{q} \widetilde{g}$, offering the prospect to reach much higher squark masses, perhaps exploiting jet substructure techniques due to the large boost of the gluino \cite{Fan:2011jc}. The first study of this associated production process at 100 TeV suggests that it could probe squark masses up to about 30 TeV \cite{Ellis:2015xba}. Above this scale, we have no direct access to squarks of any flavor, and gluinos become the most sensitive indirect probe of squark properties.

\begin{figure}[tp]
  \begin{center}
\includegraphics[width=0.65\textwidth]{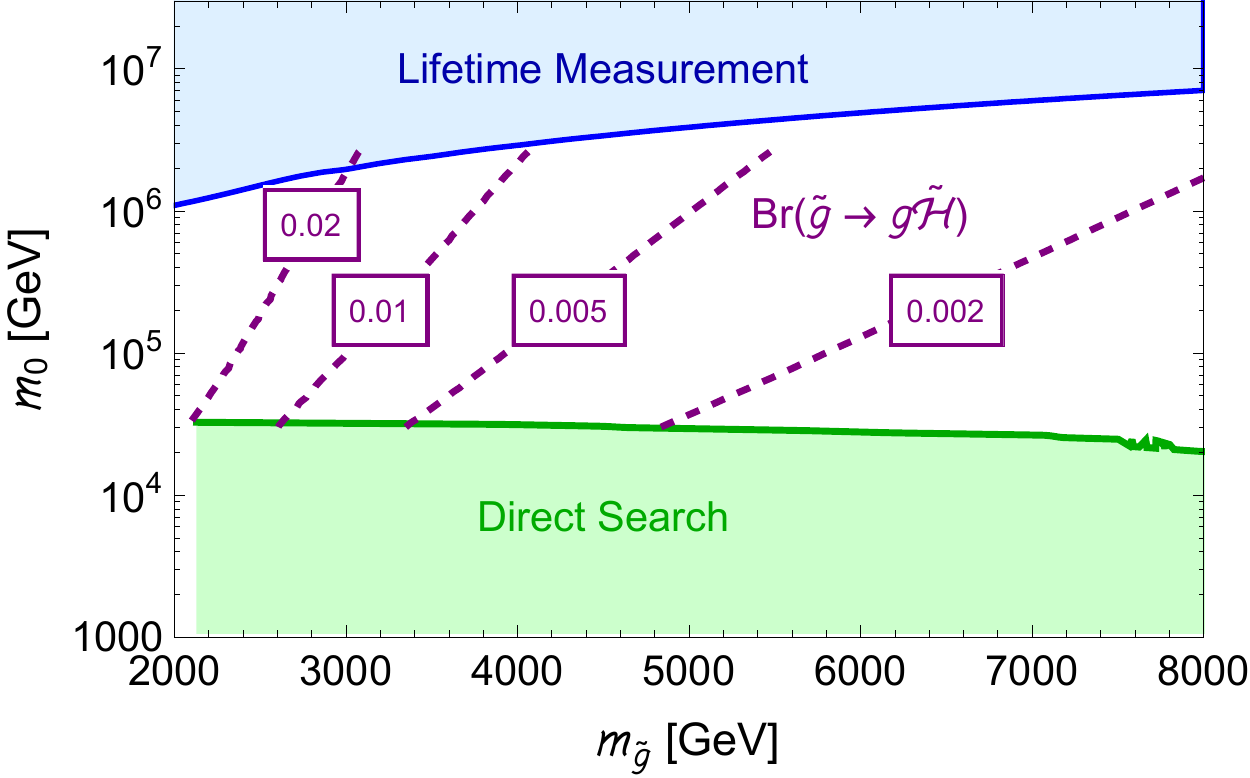}
\end{center}
\caption{The gluino--squark mass plane, categorized by means of experimentally probing the squark mass scale. At low squark masses (green region), associated squark-gluino production offers direct access to valence squark masses \cite{Ellis:2015xba}. At large squark masses (blue region), a displaced gluino vertex could be measured, as the lifetime is above 100 microns \cite{ArkaniHamed:2004fb,Arvanitaki:2012ps,ArkaniHamed:2012gw}. The intermediate region is more challenging: here the two-body gluino decay $\widetilde{g} \to g \widetilde{H}$, with branching ratio indicated on purple dashed contours, is logarithmically sensitive to the scalar mass scale \cite{Sato:2012xf}. For concreteness we have plotted the branching ratio for the choice of $\tan \beta$ that achieves a 125 GeV higgs mass for given $m_0$ and fixed $\mu = 200$ GeV, $m_1 = 700$ GeV and $m_2 = 1$ TeV. 
}
\label{fig:gluinoscalarplane}
\end{figure}%

Interestingly, gluinos decays can also yield information about
$\tan\beta$. We will assume that $M_3 > M_1,
M_2, |\mu|$, so that the gluino can decay to all neutralinos and
charginos. This is a typical spectrum obtained in many models.
Any of the neutralinos and charginos will cascade promptly to the LSP. In our studies below, we will also assume that the mass scales $M_1, M_2, M_3$, and $|\mu|$ have been accurately measured, either in direct electroweakino production processes or in cascades through the gluino. Such mass measurement problems are well-studied (see e.g.~\cite{Barr:2010zj,Barr:2011xt,Agrawal:2013uka,Debnath:2016gwz} for some entry points to the literature), so we believe this assumption to be reasonable.

\subsection{Scalar mass measurement} 
\label{sec:scalarmassmeasure}

As mentioned above, we focus on scalar mass scales of tens of TeV, 
beyond the direct reach of a 100 TeV collider.  For very large scalar
masses, the lifetime of the gluino becomes long
enough to measure: for a 2 TeV gluino, scalar mass scales $m_0 \sim 1000$
TeV result in a 100 micron lifetime \cite{ArkaniHamed:2004fb,Arvanitaki:2012ps,ArkaniHamed:2012gw}. This reach can be extended to lower scalar masses with improved detector technology, but since the
lifetime depends on the fourth power of the scalar mass, dramatic
improvement is unlikely. We see that in the region of scalar masses
$30\TeV \lesssim m_0 \lesssim 1000\TeV$, we can rely on neither direct squark production or gluino lifetime observations, and only have access to gluino branching fractions.

\begin{figure}[tp]
  \centering
  \includegraphics{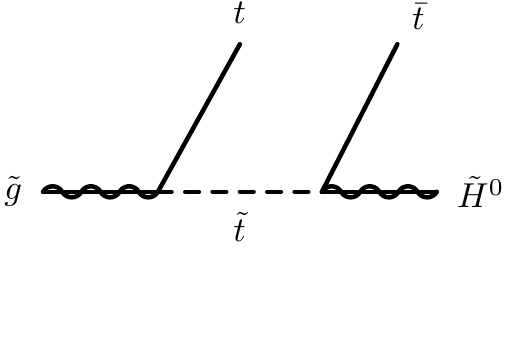}
  \qquad
  \includegraphics{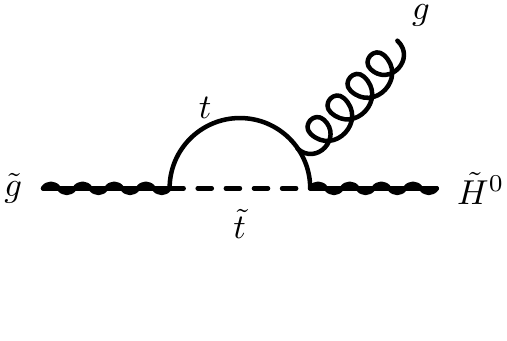}
  \vspace{-10mm}
  \caption{Tree-level and one-loop decays of the gluino.}
  \label{fig:go-decay}
\end{figure}

Gluino decays arise from dimension-six operators generated by
integrating out squarks. The tree-level decays of gluinos all have a similar dependence on the
scalar mass scale, and hence ratios of these decay widths are not
sensitive to the overall scalar mass scale.
However, the gluino decay to a gluon and a neutral higgsino, $\tg \to
g \tH^0_{1,2}$, proceeds at one loop and picks up logarithmic
contributions from scales between the scalar mass scale and the top
mass \cite{Toharia:2005gm,Gambino:2005eh}. Thus, this partial width has an 
additional {\em
logarithmic} sensitivity to the scalar mass scale. (Note that the one loop decay to a gluon and a bino does not have the same logarithmic enhancement.)
The gluino branching ratio to
gluon plus higgsino has been discussed as a key probe in this region
\cite{Sato:2012xf}. The parameter space and the
possible probes are summarized in Figure~\ref{fig:gluinoscalarplane}.
We show an example
tree-level and one-loop decay of the gluino in
Figure~\ref{fig:go-decay}.

The following ratio of two- to three-body decays is a clean probe of
the scalar mass scale~\cite{Toharia:2005gm}:
\begin{align}
  \frac{\Gamma(\tg \to g \tH^0)}
  {\Gamma(\tg \to t{\bar t} \tH^0)}
  \propto 
  \frac{m_t^2}{m_\tg^2} 
  \log^2 \frac{m_\tildet^2}{m_t^2}.
  \label{eq:gluinoloopratio}
\end{align} 
The decay widths here are summed over the two neutral higgsino final
states, since they can be difficult to distinguish from one
another experimentally. For very large values of $m_{\widetilde{t}}$, the
logarithm becomes large and resummation is required for accurate
predictions~\cite{Gambino:2005eh}. This tends to
flatten out the scalar mass
dependence, but in any case it is a small effect for the values of
$m_{\widetilde{t}}$ we are interested in.
Note that since the same particles and couplings appear in the two
diagrams, the ratio is relatively insensitive to the details of the
scalar mass spectrum, or to the value of $\tan\beta$.

\begin{figure}[t]\begin{center}
\includegraphics[width=0.47\textwidth]{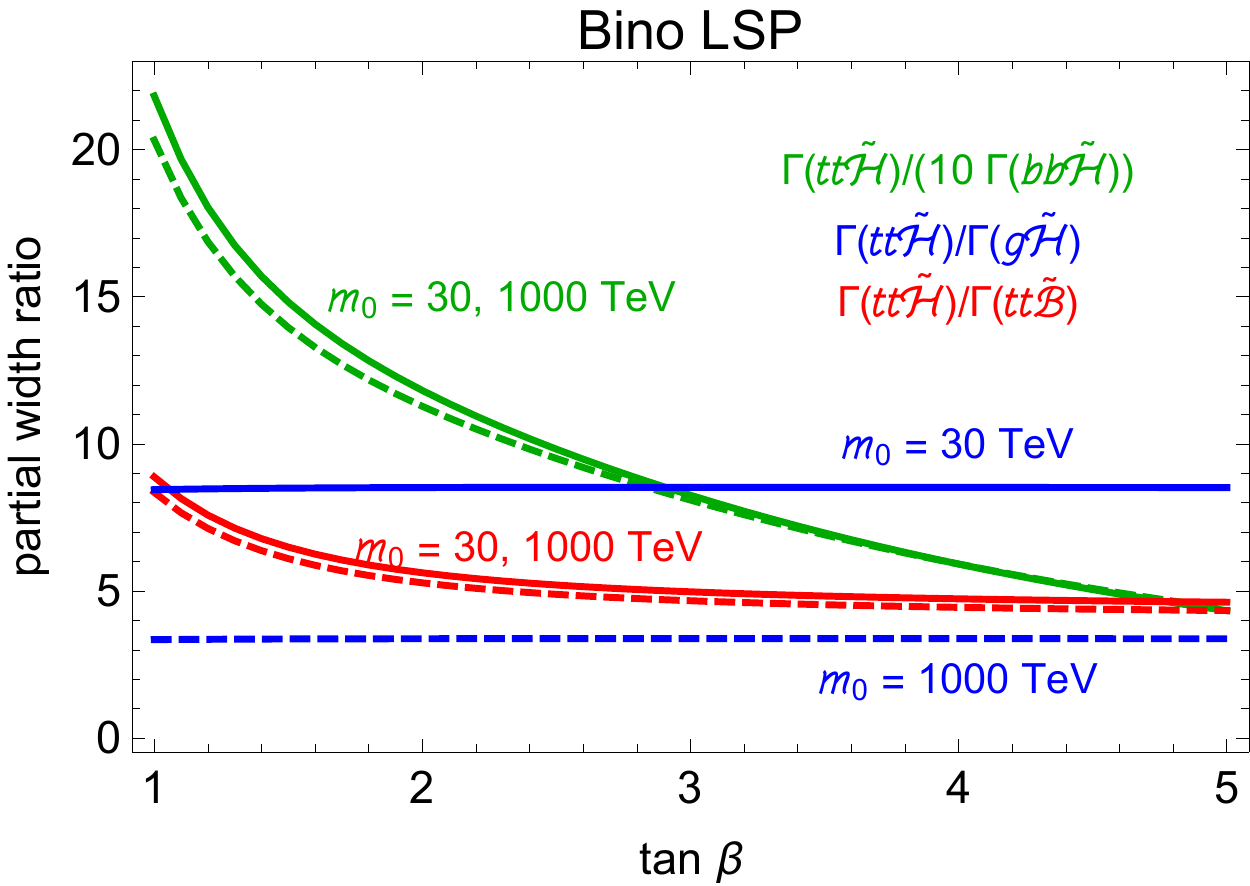}~~\includegraphics[width=0.49\textwidth]{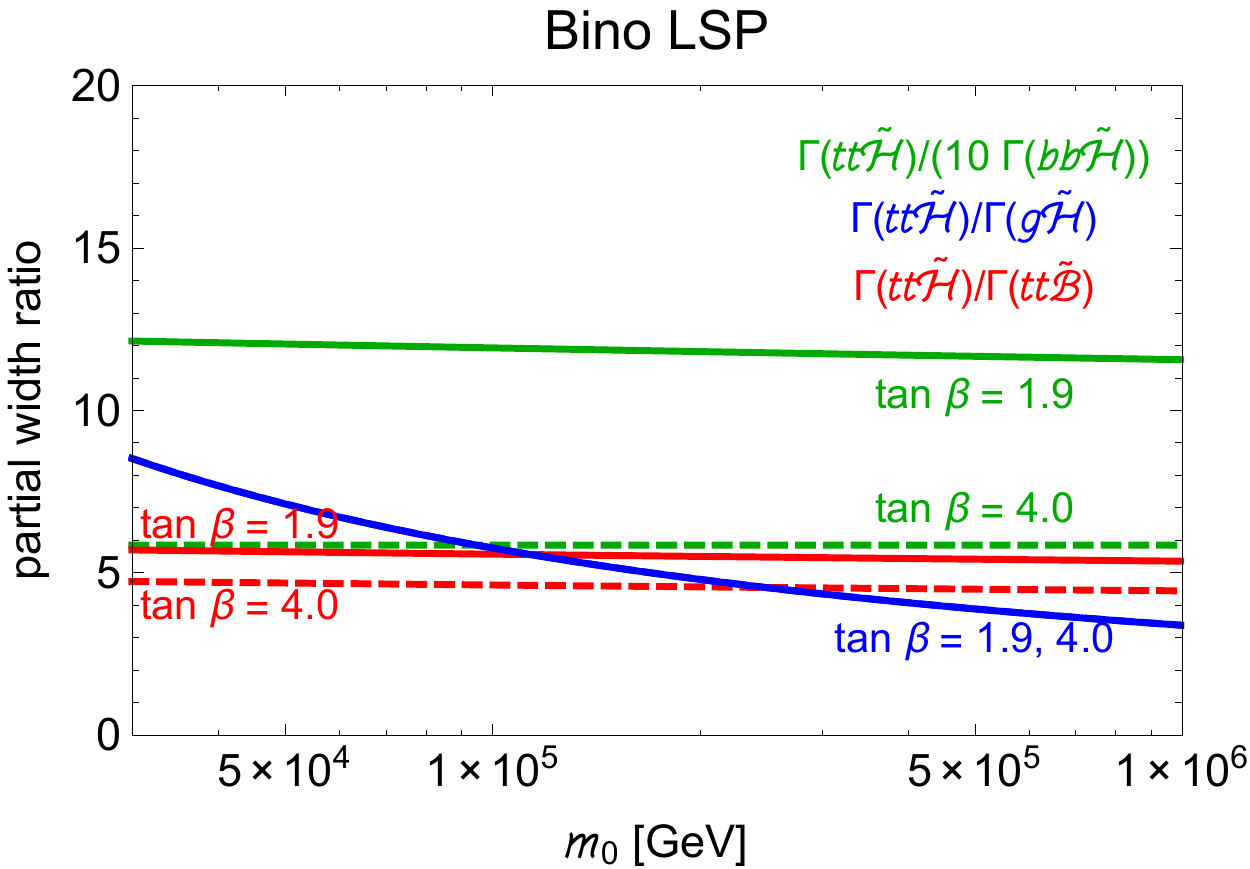}\\
\includegraphics[width=0.47\textwidth]{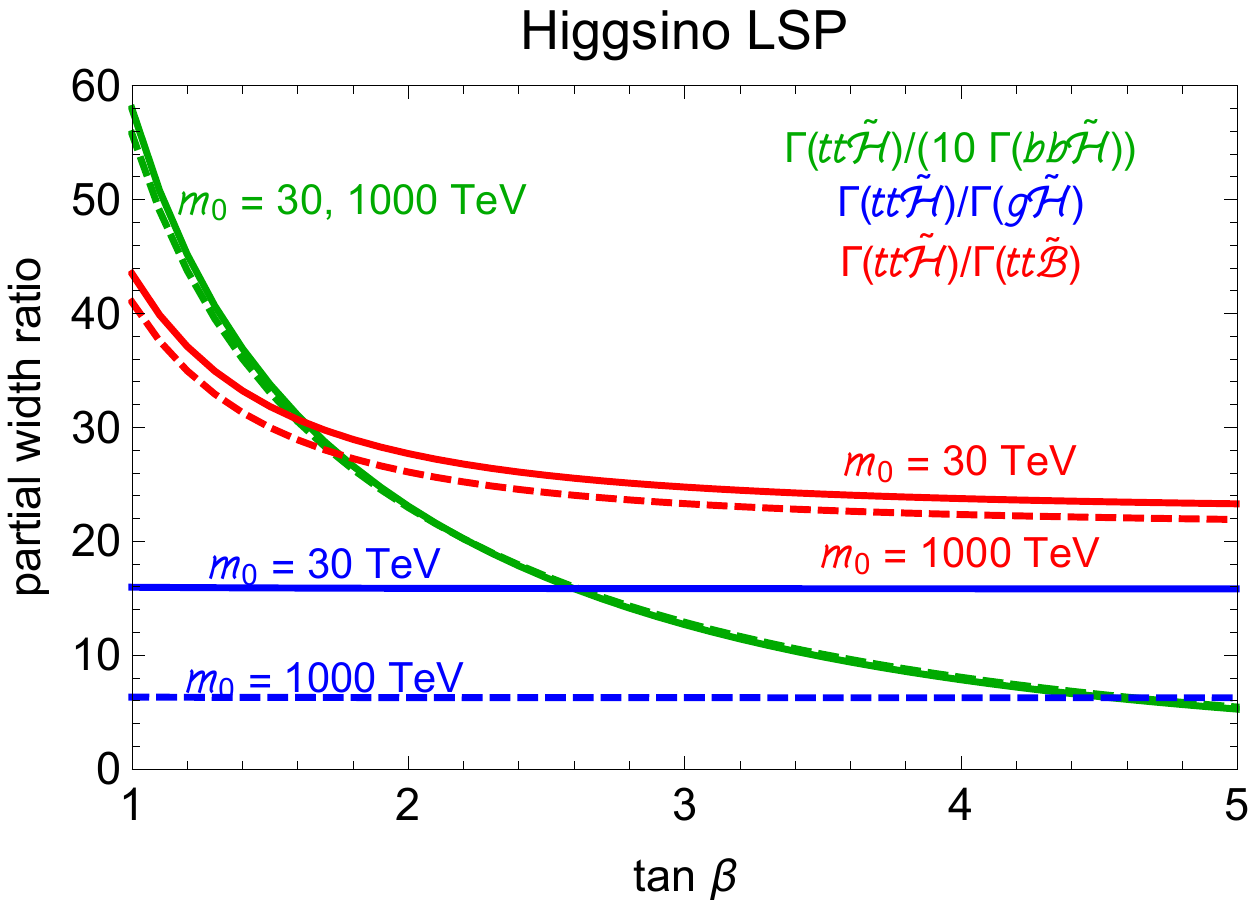}~~\includegraphics[width=0.49\textwidth]{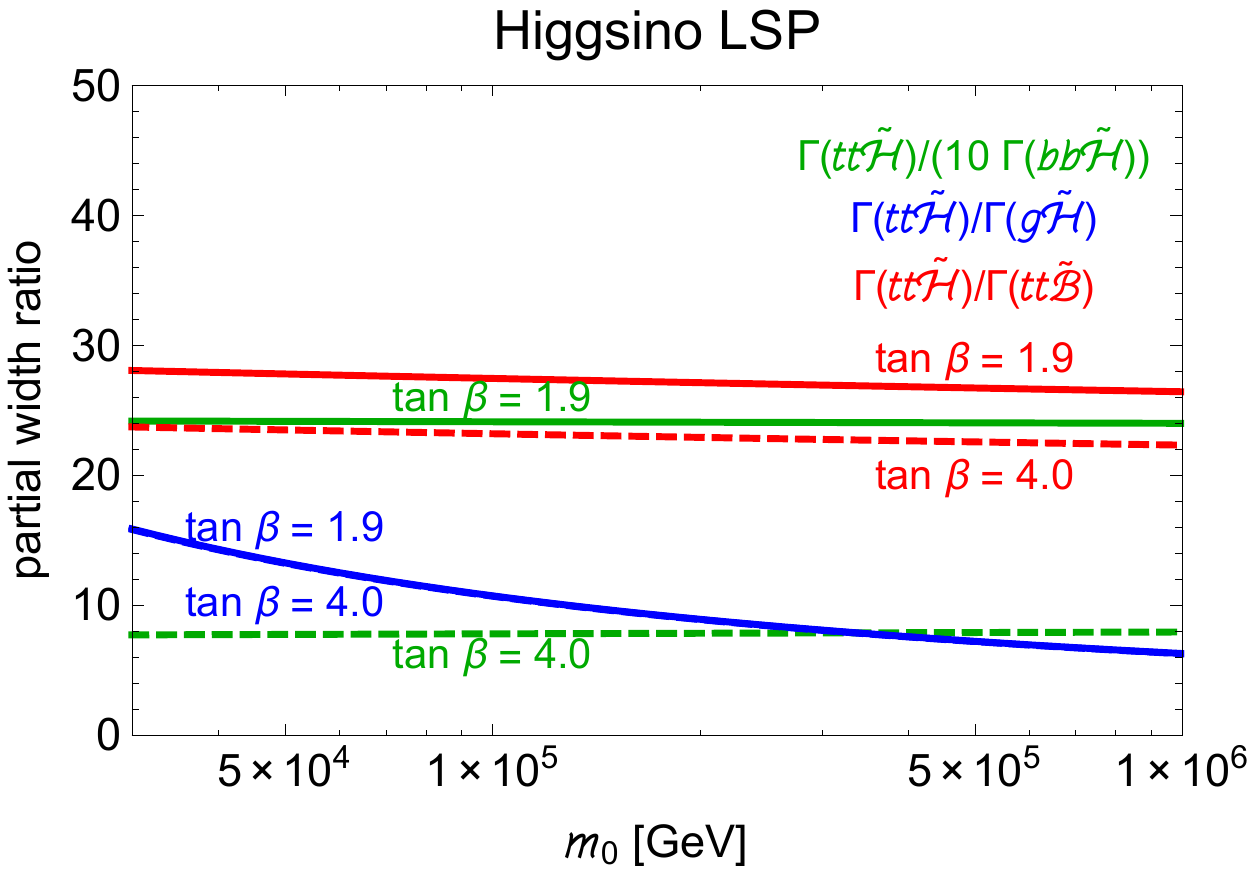}
\end{center}
\caption{Gluino branching ratios, plotted as functions of a universal scalar mass scale $m_0$ and $\tan \beta$. We choose two benchmark points, both with $M_3 = 2$ TeV: in the top row, $M_1 = 200$ GeV, $M_2 = 400$ GeV, and $\mu = 800$ GeV; in the bottom row, $M_1 = 700$ GeV, $M_2 = 1$ TeV, and $\mu = 200$ GeV. Because the $b\overline{b}$ width is very small, we have multiplied it by 10 to make the green curves visible in the plot.}
\label{fig:gluinobranchingplots}
\end{figure}%

\subsection{Gluino decays and tan \texorpdfstring{$\bm{\beta}$}{$\beta$}}

Gluino decays to higgsinos have a dependence on $\tan\beta$ due to
the appearance of the
Yukawa couplings $Y_u \propto
1/\sin \beta$ and $Y_d \propto 1/\cos \beta$. Thus there are a number
of options to measure ratios of decay rates to measure $\tan\beta$.

We can measure the rate of a gluino decay to higgsino
relative to the rate to gauginos:
\begin{equation}
\frac{\Gamma(\tg \to t{\bar t} \tH^0)}{\Gamma(\tg \to t{\bar t} \tB^0)}, \frac{\Gamma(\tg \to t{\bar t} \tH^0)}{\Gamma(\tg \to t{\bar t} \tW^0)} \propto \frac{1}{\sin^2 \beta}. \label{eq:gluinotreeratio} 
\end{equation}
Decays to binos and winos can also help resolve additional parameters,
such as the 
left- and right-handed stop masses.
Note that the
dependence on $\tan \beta$ is mild over the range we are 
interested in and would need very small systematic uncertainties in
efficiencies at colliders (<5\%) to be useful.

The decays to $b$-quarks and a higgsino have a  steeper $\tan \beta$
dependence. In
particular, if we can measure the ratio between two decays to
higgsinos, we can obtain
\begin{equation}
\frac{\Gamma(\tg \to b{\bar b} \tH^0)}{\Gamma(\tg \to t{\bar t} \tH^0)} \propto \tan^2 \beta. \label{eq:gluinobbbarratio}
\end{equation}
The decay rate in the numerator is very small
for the $\tan \beta$ values we are interested in
due to the small $b$-Yukawa. Another possible
 measurement is the ratio $\Gamma(\tg \to b {\bar b}
\tH^0)/\Gamma(\tg \to g \tH^0)$. This has the same $\tan \beta$
dependence as above, is a larger ratio, and the events being compared may be
more similar kinematically. The denominator is sensitive to the scalar
mass scale, so to measure $\tan\beta$ we have to separately measure
the $m_0$ dependence as well.

In figure \ref{fig:gluinobranchingplots}, we plot the 
observable $\Gamma(t{\bar t}\tH^0)/\Gamma(g \tH^0)$ (blue) which is
sensitive to the scalar mass scale. We also show $\tan\beta$-dependent
observables
$\Gamma(t{\bar t}\tH^0)/\Gamma(t{\bar t}\tB^0)$ (red) 
and
$\Gamma(t{\bar t}\tH^0)/\Gamma(b {\bar b} \tH^0)$ (green). The latter has a
much steeper dependence on $\tan\beta$, but is small; hence, the curve has
been rescaled by a factor of 10 to fit in the plot. All decay rates
include  resummation effects.
The latter two observables (in green and red) are also mildly 
sensitive to  the scalar mass scale due to renormalization group
mixing among the different
dimension-six operators.

\section{Electroweak observables sensitive to tan \texorpdfstring{$\bm{\beta}$}{$\beta$}}
\label{sec:ewkobservables}

In addition to gluino branching ratios, the electroweak sector can serve as a probe of $\tan \beta$. Because we work in the limit where the heavy higgs bosons are decoupled, we can study characteristics of the light higgs boson or of the electroweakinos. We will see that some electroweakino branching ratios depend dramatically on $\tan \beta$, vanishing when $\tan \beta \to 1$. Other branching ratios, including $h \to \gamma\gamma$, are sensitive to $\tan \beta$ in more subtle ways. Dark matter direct detection can also provide a probe of $\tan \beta$ by measuring neutralino couplings to the $Z$ and $h$ bosons.

\subsection{Blind spot at tan \texorpdfstring{$\bm{\beta}$}{$\beta$} = 1}
\label{subsec:blindspot}

A number of observables that are sensitive to $\tan\beta$ can be
understood as arising from the ``blind spot'' at $\tan\beta=1$.
The central point is that there is an enhanced parity symmetry
at $\tan\beta=1$ which restricts various observables. Hence
deviations from $\tan\beta=1$ are reflected in deviations from these
restrictions.

Higgsinos come from two doublets of equal
and opposite hypercharge, 
\begin{align} 
  \tH_u \equiv 
  \begin{pmatrix} 
    \tH_u^+ \\
    \tH^0_u 
  \end{pmatrix} 
  \in 
  {\mathbf 2}_{+1/2}, \quad \tH_d 
  \equiv
  \begin{pmatrix} 
    \tH^0_d \\ \tH^-_d 
  \end{pmatrix}
  \in
  {\mathbf 2}_{-1/2}.  
\end{align} 
It is useful to define the basis $\tH_\pm^0$,
\begin{align} 
  \tH_{\pm}^0 &=
  \frac{1}{\sqrt{2}} \left(\tH_u^0 \pm \tH_d^0 \right)
  \,.  
\end{align} 
The $\mu$ term gives rise to a Dirac mass which may be thought of as equal and opposite Majorana masses for $\tH^0_+$ and $\tH^0_-$. Mixing with the bino and wino splits the two Majorana mass eigenstates, but they remain approximately $\tH^0_\pm$.

Expanding out the kinetic terms, we find that the $Z$ boson coupling to
the neutral higgsinos is off-diagonal in the $\tH^0_\pm$ basis:
\begin{align} 
  i \tH_u^\dagger {\overline \sigma}^\mu D_\mu
  \tH_u + i \tH_d^\dagger {\overline \sigma}^\mu D_\mu \tH_d
  &\supset \frac{g}{2 \cos\theta_W} Z_\mu \left(\tH_u^{0\dagger} \sigmabar^\mu \tH^0_u - \tH_d^{0\dagger} \sigmabar^\mu \tH^0_d\right)
  \nonumber \\
& =  \frac{g}{2 \cos \theta_W} Z_\mu
  \left(\tH_+^{0\dagger} {\overline \sigma}^\mu \tH_-^0 +
  \tH_-^{0 \dagger} {\overline \sigma}^\mu \tH^0_+ 
  \right).  
\end{align} 
The supersymmetric counterparts to these terms are the gauge-Yukawa couplings involving neutralinos,
\begin{align}  \label{eq:susyYukawas}
  \mathcal{L}
  & \supset \frac{1}{\sqrt{2}} \left(g \tW^0 - g' \tB^0 \right)\left(H_u^{0\dagger} \tH_u^0 - H_d^{0\dagger} \tH_d^0\right) + \mathrm{h.c.} \nonumber \\
  &\rightarrow \frac{\cos\beta}{2\sqrt{2}} (v+h) \left(g \tW^0 - g' \tB^0\right) \left[(1 - \tan \beta) \tH_+^0 - (1 + \tan \beta) \tH_-^0\right] + \mathrm{h.c.},
  \end{align}
where we have used the replacement
\begin{align}
H_u^0 &\to \frac{1}{\sqrt{2}} (v+h) \sin\beta
, \quad 
H_d^0 \to \frac{1}{\sqrt{2}} (v+h) \cos\beta,
\end{align}
which applies in the decoupling limit when all the other scalars are
heavy.
We see that at $\tan\beta=1$, $\tH_+^0$ does not couple to the higgs
or mix with the bino or the wino. This is a consequence of 
a parity symmetry under which $\tH_+^0$ and the $Z$ boson are odd, but
all other neutralinos and the higgs are even.
The absence of mixing also implies that at $\tan\beta=1$, 
$\tH_+^0$ is a 
mass eigenstate. 

If a small effect splits the two Majorana mass eigenstates
slightly, then $Z$-mediated physical processes are always
off-diagonal, e.g.~collider production $e^+e^- \to \tH^0_+ \tH^0_-$ or
direct detection $\tH^0_+ N \to \tH^0_- N$. Thus, if $\tH^0_+$ is the
dark matter, $Z$-mediated direct detection at tree-level is inelastic, suppressing the rate.
Subleading effects may
lead to a mass eigenbasis not perfectly aligned with $\tH^0_\pm$.

The neutral cascade decays of $\tB^0,\tW^0$ proceed through the
gauge-Yukawa couplings with the higgsinos. 
Due to conserved parity at $\tan\beta=1$, all such decays to or from
$\tH_+^0$ are accompanied by a $Z$ (using the mixing of the gaugino with $\tH_-^0$ and
the off-diagonal $Z$ couping). On
the other hand, decays to $\tH_-^0$ all produce higgses. Thus,
depending on the spectrum, the relative fraction of $Z$ vs.~$h$
in the final states are a diagnostic of deviation from the
$\tan\beta=1$ limit. 

For example, below we will discuss a benchmark spectrum in which higgsinos are the
NLSPs and the LSP is the bino (but could also be a wino, with little change in the physics). In that case, we find that the number of $ZZ+\tB^0 \tB^0$ events in $\tH_+^0
\tH_-^0 (\sim
\tchi_2^0 \tchi_3^0)$ production has a strong $\tan\beta$ dependence,
and hence can be used for its measurement. An alternative observable arises from $\tW^0 \to \tH^0 \to \tB^0$ cascades; in this case, we find that cascades containing both a $Z$ and an $h$ are suppressed at $\tan \beta = 1$, where (\ref{eq:susyYukawas}) implies that
\begin{equation}
\frac{\Gamma(\tW^0 \to Z h \tB^0)}{\Gamma(\tW^0 \to Z Z \tB^0)+\Gamma(\tW^0 \to h h \tB^0)} \propto \left(\frac{1 - \tan \beta}{1 + \tan \beta}\right)^2.
\end{equation}

\subsection{Higgsino LSPs}
\label{subsec:higgsinoLSP}

In the case of higgsino LSPs, the heavier higgsinos decay promptly to the lightest higgsino mass eigenstate. Mass splittings within the higgsino multiplet are small, so the decay products from these transitions are soft and difficult to detect. The heavier gauginos decay promptly to higgsinos through the supersymmetric gauge interactions. We can see from (\ref{eq:susyYukawas}) that in principle these decays carry $\tan \beta$ information---for instance, $\tW^0 \to h \tH^0_+$ turns off at $\tan \beta = 1$---but because the different higgsino mass eigenstates are nearly indistinguishable experimentally, it is difficult to use this information. On the other hand, if we can find events (perhaps in cascades starting with wino or gluino pair production) containing the decay $\tH^0_2 \to Z^* \tH^0_1 \to \ell^+ \ell^- \tH^0_1$ and measure the dilepton mass spectrum, we can measure the higgsino mass difference, which depends on $\tan \beta$. The leading approximation to the neutral higgsino mass splitting is $\tan \beta$ independent and scales as $m_Z^2/M_{1,2}$, so the effect arises only from a smaller term of order $\mu m_Z^2/M_{1,2}^2 \sin(2\beta)$ (see e.g.~\cite{Martin:1997ns}). There is also a small effect of $\tan \beta$ on the fraction of events containing such a $\tH^0_2 \to Z^* \tH^0_1$ transition. 

We will return to the case of higgsino LSPs below in \S\ref{subsec:darkmatter}, where we will see that complementary information from dark matter direct detection experiments may help to pin down $\tan \beta$.

\subsection{Higgsinos heavier than gauginos}
\label{sec:tanbetahiggsinosheavy}

The $\tan \beta$ sensitivity we have discussed so far is associated with the higgsino sector. If we have a spectrum with $\mu > M_2 > M_1$, these results are difficult to apply, because the higgsino pair production cross section is much smaller than the wino production rate. However, it may be possible to measure $\tan \beta$ through the relative size of $\tW^0 \to h \tB^0$ and $\tW^0 \to Z \tB^0$ decays. Integrating out the higgsino, we find effective wino--bino couplings from 
\begin{equation}
{\cal L}_{\rm eff} \supset \frac{g g'}{\mu} \tB \tW^i H_u \cdot T^i H_d + \frac{gg'}{2\mu^2} \tB \sigmabar^\mu \tW^{i\dagger} \left(H_d^\dagger i \overset{\leftrightarrow}{D}_\mu \sigma^i H_d - H_u^\dagger i\overset{\leftrightarrow}{D}_\mu T^i H_u\right) + {\rm h.c.}
\end{equation}
The first term allows only the decay $\tW^0 \to h \tB$; the second, $\tW^0 \to Z \tB$. The former decay arises from an operator containing both $H_u$ and $H_d$ and so is suppressed at large $\tan \beta$. In the limit $\mu \gg M_2 \gg m_h$ at fixed $M_1/M_2$, the ratio of decay widths is
\begin{equation}
\frac{\Gamma(\tW^0 \to h \tB^0)}{\Gamma(\tW^0 \to Z \tB^0)} \approx \frac{16 \tan^2\beta}{(1-\tan^2\beta)^2}\frac{\mu^2}{M_2^2} \left(\frac{1+M_1/M_2}{1-M_1/M_2}\right)^2. \label{eq:winobinoratio} 
\end{equation}
This could be an interesting observable for $\tan \beta$ measurement. Notice that to make use of it we must measure the mass scale $\mu$, either through direct production of higgsinos or through gluino decays to higgsinos. In the case $\mu > M_1 > M_2$, similar reasoning applies but we do not directly produce binos, so the $\tB^0 \to \tW^0$ branching fractions could be measured only if we produce the bino from a heavier particle like the gluino.

\subsection{Higgs boson branching ratios}

\begin{figure}[!h]\begin{center}
\includegraphics[width=0.95\textwidth]{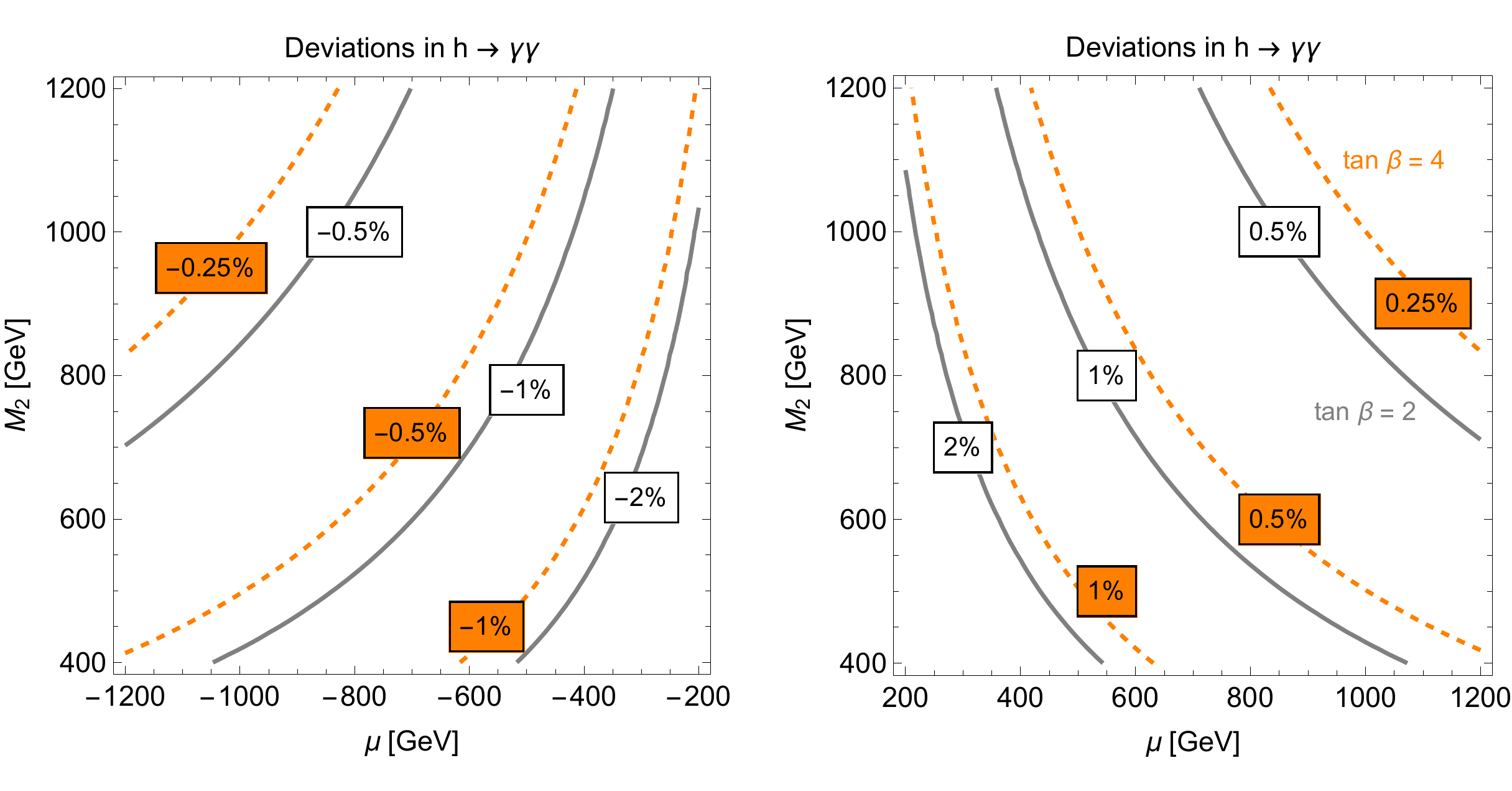}
\end{center}
\caption{Effect of chargino loops on the $h \to \gamma\gamma$ branching fractions, which can lead to percent-level deviations from the Standard Model expectation. Left: $\mu < 0$; right: $\mu > 0$. Solid curves: $\tan \beta = 2$. Dashed orange curves: $\tan \beta = 4$. }
\label{fig:hgammagammaplot}
\end{figure}%

In the MSSM, higgs boson properties may be modified by a variety of effects, including mixing with the heavy higgs bosons. However, in the split SUSY limit, only the Standard Model-like higgs boson is light, and most of these effects decouple. In this case the leading deviations in higgs properties arise from loops of electroweakinos \cite{Diaz:2004qt}. The most detectable of these effects is the modification to the $h \to \gamma\gamma$ decay that already arise at one-loop order in the Standard Model. For the decay to two photons, the modification of the partial width is readily computed from the low-energy theorem \cite{Ellis:1975ap,Shifman:1979eb}: 
\begin{equation}
\frac{\Gamma(h \to \gamma\gamma)}{\Gamma(h \to \gamma \gamma)_{\rm SM}} \approx 1 + \frac{0.82 m_W^2 \sin(2\beta)}{\mu M_2 - m_W^2 \sin(2\beta)}.
\end{equation}
The deviation is largest at small values of $\tan \beta$. Since $|\mu M_2| > m_W^2 \sin(2\beta)$, the sign of the deviation depends on that of $\mu M_2$: fixing $M_2$ to be positive, when $\mu$ is positive, $\Gamma(h \to \gamma\gamma)$ is enhanced due to a constructive interference between the electroweakino loop and the Standard Model $W$ loop; when $\mu$ is negative, $\Gamma(h \to \gamma\gamma)$ is reduced due to a destructive interference. Thus measuring a deviation in the $h\gamma\gamma$ branching fraction not only gives us a clue about $\tan\beta$ but also the sign of $\mu$.
We have illustrated this effect in Figure~\ref{fig:hgammagammaplot}. The effect is small: only a 2\% increase (decrease) in the branching ratio in the optimistic case $\tan \beta \approx 2$ for the point $|\mu| \approx M_2 \approx 500~{\rm GeV}$. The expected precision of the $h\gamma\gamma$ coupling measurement at future $e^+ e^-$ colliders will not be sensitive to such small deviations: for example, FCC-ee would achieve about a 1.5\% measurement of the {\em coupling} (and thus a 3\% sensitivity to the branching fraction) \cite{Gomez-Ceballos:2013zzn}. However, hadron colliders offer a unique opportunity to measure the {\em ratio} of photon and $Z$ branching fractions \cite{ATLAS:2013hta,Peskin:2013xra}. Systematic uncertainties that plague the measurement of individual couplings, for instance in luminosity or parton distribution functions, cancel in the ratio $\Gamma(h \to \gamma\gamma)/\Gamma(h \to Z Z^*)$. At FCC-hh, the very large luminosity and higgs production rates could offer the possibility of sub-percent-level {\em statistical} uncertainties on such ratios, even when making an additional selection cut on the higgs $p_T$ to boost the signal-to-background ratio \cite{Contino:2016spe}. It remains to be seen how well systematics could be controlled, but there is at least the prospect that precision higgs measurements could allow us to indirectly infer the value of $\tan \beta$, at least in a portion of the $(M_2, \mu)$ plane. 
Lastly, $\Gamma(h \to Z \gamma)$ could also be modified by a light electroweakino loop in a similar way. Yet it is more difficult to measure the $Z\gamma$ branching fraction precisely and we will not pursue it here.

\subsection{Charged wino lifetime}
\label{subsec:chargedwino}

\begin{figure}[!h]\begin{center}
\includegraphics[width=0.65 \textwidth]{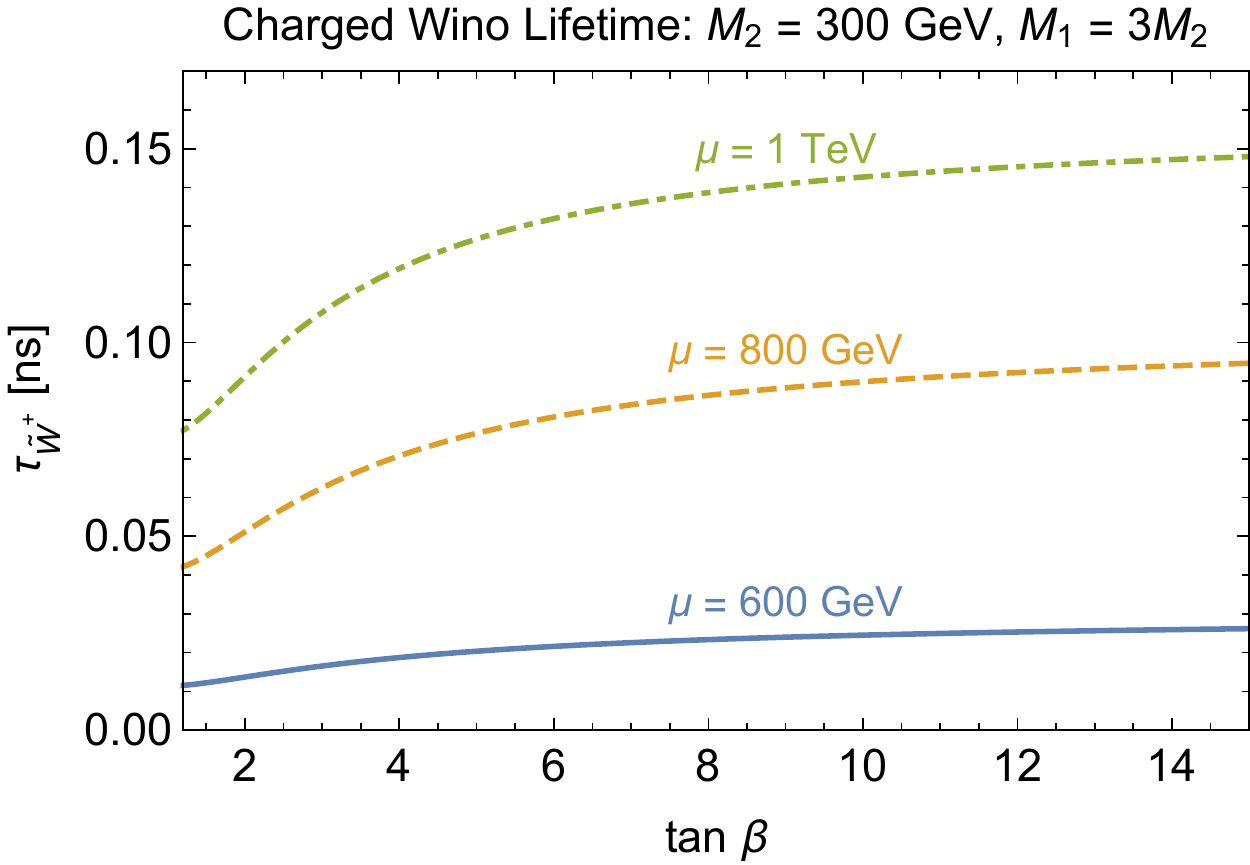}
\end{center}
\caption{Dependence of the charged wino lifetime on $\tan \beta$ in a scenario with mostly-wino LSP. We take $M_1 = 3 M_2$, as in AMSB, and fix $M_2 = 300~{\rm GeV}$. We have plotted three different choices of $\mu$. The lifetime is typically in the range of $0.1~{\rm ns}$, but varies by an order one amount as $\tan \beta$ varies.}
\label{fig:chargedwinolifetime}
\end{figure}%

The charged and neutral wino states are nearly degenerate; when the wino is the LSP, we can exploit this degeneracy for a lifetime measurement. 
The tree-level splitting between charged and neutral winos is approximately given by \cite{Gherghetta:1999sw}
\begin{align} \label{eq:winotreesplitting}
\delta m^{\rm tree}_{\tW} \approx \frac{m_W^4 \sin^2(2\beta)}{(M_1 - M_2) \mu^2} \tan^2 \theta_W + 2 \frac{m_W^4 M_2 \sin(2\beta)}{(M_1 - M_2)\mu^3} \tan^2 \theta_W + \frac{m_W^4 M_2}{2\mu^4} + \ldots
\end{align}
Notice that the first two terms vanish as $\tan \beta \to \infty$, while the third term remains finite---but goes to zero more quickly when the higgsino is decoupled. This third, $M_1$-independent piece of the mass splitting in \eqref{eq:winotreesplitting} arises from a dimension-six kinetic correction $\propto \frac{1}{\mu^2} \iu \epsilon_{ijk} (h^\dagger \sigma^i h) \tW^{j\dagger} \sigmabar^\mu D_\mu \tW^k$ generated by integrating out the higgsinos. At leading order this gives equal and opposite wavefunction renormalization corrections to $\tW^+$ and $\tW^-$, leading to no shift in the chargino mass, but at second order it gives a mass shift proportional to $1/\mu^4$. The $M_1$-dependent mass differences arise from higher dimension operators like $(h^\dagger \sigma^i h)(h^\dagger \sigma^j h) \tW^i \tW^j$. Beyond these tree-level effects, there is a loop correction even in the pure wino case,
\begin{align} \label{eq:winoloopsplitting}
\delta m^{\rm loop}_{\tW} \approx \frac{\alpha m_W}{2(1 + \cos \theta_W)} \approx 165~{\rm MeV},
\end{align}
which is known to two-loop order \cite{Ibe:2012sx}. For a given point in parameter space, we compute the tree-level mass splitting by diagonalizing the full mass matrices, then add the loop correction, and finally infer the lifetime from formulas in ref.~\cite{Ibe:2012sx}.

These small mass splittings lead to a ``disappearing track'' signal at colliders, due to the relatively long lifetime of the charged wino \cite{Chen:1995yu,Chen:1999yf,Feng:1999fu}, which has already led to nontrivial constraints on winos at the LHC \cite{Aad:2013yna,CMS:2014gxa}. When we consider not just pure winos but the full $(M_1, M_2, \mu, \tan \beta)$ electroweakino parameter space, this constraint is stronger at large $\tan \beta$, due to the smaller tree-level splitting (\ref{eq:winotreesplitting}). Further details of the current experimental status, reinterpreted in the case of winos mixing with higgsinos and binos, may be found in \cite{Han:2016qtc,Nakai:2016atk}. We have illustrated the $\tan \beta$ dependence of the lifetime in Fig.~\ref{fig:directdetectionSDvsSI}. Increasing $\tan \beta$ from 2 to 4 increases the charged wino lifetime by 30\% to 40\%. For the limit of very pure winos for which the loop-induced splitting \eqref{eq:winoloopsplitting} dominates, it is known that a future hadron collider could discover winos via their disappearing track signature over a large part of parameter space \cite{Low:2014cba}. To use the signal as a $\tan \beta$ probe, we must work {\em away} from the pure wino limit, where $\mu$ is not too large. The higgsino and bino masses must be measured (either in electroweak production or in gluino cascade decays), and a chargino lifetime in the centimeter range must be measured precisely. This is a well-motivated and interesting challenge for studies of the tracking capabilities of future hadron colliders. The disappearing tracks may also be searched for in gluino decays \cite{Kane:2012aa}, which give the chargino an additional boost and hence a longer lifetime, perhaps making the signal more tractable.

\subsection{Dark matter direct detection}
\label{subsec:darkmatter}

In the case that electroweakinos are a mixture of higgsino and gaugino, they may be directly detected from searches for nuclear recoils mediated by higgs bosons and $Z$ bosons. The lightest neutralino's couplings to the higgs and $Z$ take the form
\be
\left[\frac{1}{2} c_{h\chi\chi} h \chi \chi + {\rm h.c.}\right]+ c_{Z\chi\chi} \chi^\dagger \sigmabar^\mu \chi Z_\mu,
\ee
where \cite{Haber:1984rc,Dreiner:2008tw}
\begin{align}
c_{h \chi \chi}^* &= (N_{13} \sin \alpha + N_{14} \cos \alpha) (g N_{12} - g' N_{11}) \nonumber \\
&\approx \frac{g m_W (1 + \sign(\mu) \sin(2\beta))}{2} \left[\frac{1}{M_2 - |\mu|} + \frac{\tan^2 \theta_W}{M_1 - |\mu|}\right], \label{eq:chchichi} \\
c_{Z \chi \chi} &= \frac{g}{2 \cos \theta_W} \left(|N_{14}|^2 - |N_{13}|^2\right) \nonumber \\
&\approx \frac{g m_W^2}{4 |\mu| \cos \theta_W} \cos(2\beta) \left[\frac{1}{M_2 - |\mu|} + \frac{\tan^2 \theta_W}{M_1 - |\mu|}\right]. \label{eq:cZchichi}
\end{align}
and we have provided approximations valid in the limit $M_{1,2} \gg |\mu| > 0$. Scattering through the higgs relies on the higgs--higgsino--gaugino vertices, and so requires mixing of the higgsino with the bino or wino. Scattering with the $Z$ proceeds entirely through higgsino components; however, in the pure higgsino limit, the mass eigenstates are $\tH_\pm^0$, so $c_{Z\chi\chi} \to 0$. Hence, spin-dependent scattering {\em also} requires mixing with the bino or wino. (These effects are also easily understood in terms of the higgsino effective theory arising when the bino and wino are integrated out \cite{Essig:2007az,Nagata:2014wma}.) As signaled by the factor of $\cos(2\beta)$ in \eqref{eq:cZchichi}, searches for spin-dependent scattering have a ``blind spot'' at $\tan \beta = 1$, where again the eigenstates are $\tH_\pm^0$ \cite{Cohen:2010gj,Cheung:2012qy}. Hence, spin-independent and spin-dependent dark matter scattering probe similar underlying physics, but the {\em relative} rate of spin-dependent scattering can serve as a probe of $\tan \beta$.

When $\mu > 0$, the light higgsino mass eigenstate is approximately $\tH_-^0$, which couples to the higgs boson even when $\tan \beta = 1$. On the other hand, when $\mu < 0$, the light higgsino mass eigenstate is approximately $\tH^0_+$, which does not couple to the higgs when $\tan \beta \to 1$. For this reason, the spin-independent scattering rate is much smaller for negative values of $\mu$ than for positive ones.

Based on these couplings, the expected scattering rate of dark matter on a nucleon is
\begin{align} \label{eq:directrates}
\sigma_{\rm SI} &= \left|c_{h \chi \chi}\right|^2 \times (5.3 \times 10^{-43}~{\rm cm}^2), \nonumber \\
\sigma_{{\rm SD},p} &= \left|c_{Z \chi \chi}\right|^2 \times (2.9 \times 10^{-37}~{\rm cm}^2), \nonumber \\
\sigma_{{\rm SD},n} &= \left|c_{Z \chi \chi}\right|^2 \times (2.2 \times 10^{-37}~{\rm cm}^2).
\end{align}
We have taken these results from \cite{Basirnia:2016szw} (adjusting factors of 2 for conventions), which uses a recent averaging of nuclear matrix element determinations from \cite{Belanger:2013oya}. The higgs-dependent scattering rate has a $\sim 10\%$ theoretical uncertainty from our limited knowledge of the matrix element $\langle N| s {\bar s} | N\rangle$ where $|N\rangle$ is a nucleon state, but this uncertainty can be reduced in the future by further lattice QCD calculations.

\begin{figure}[!h]\begin{center}
\includegraphics[width=0.95 \textwidth]{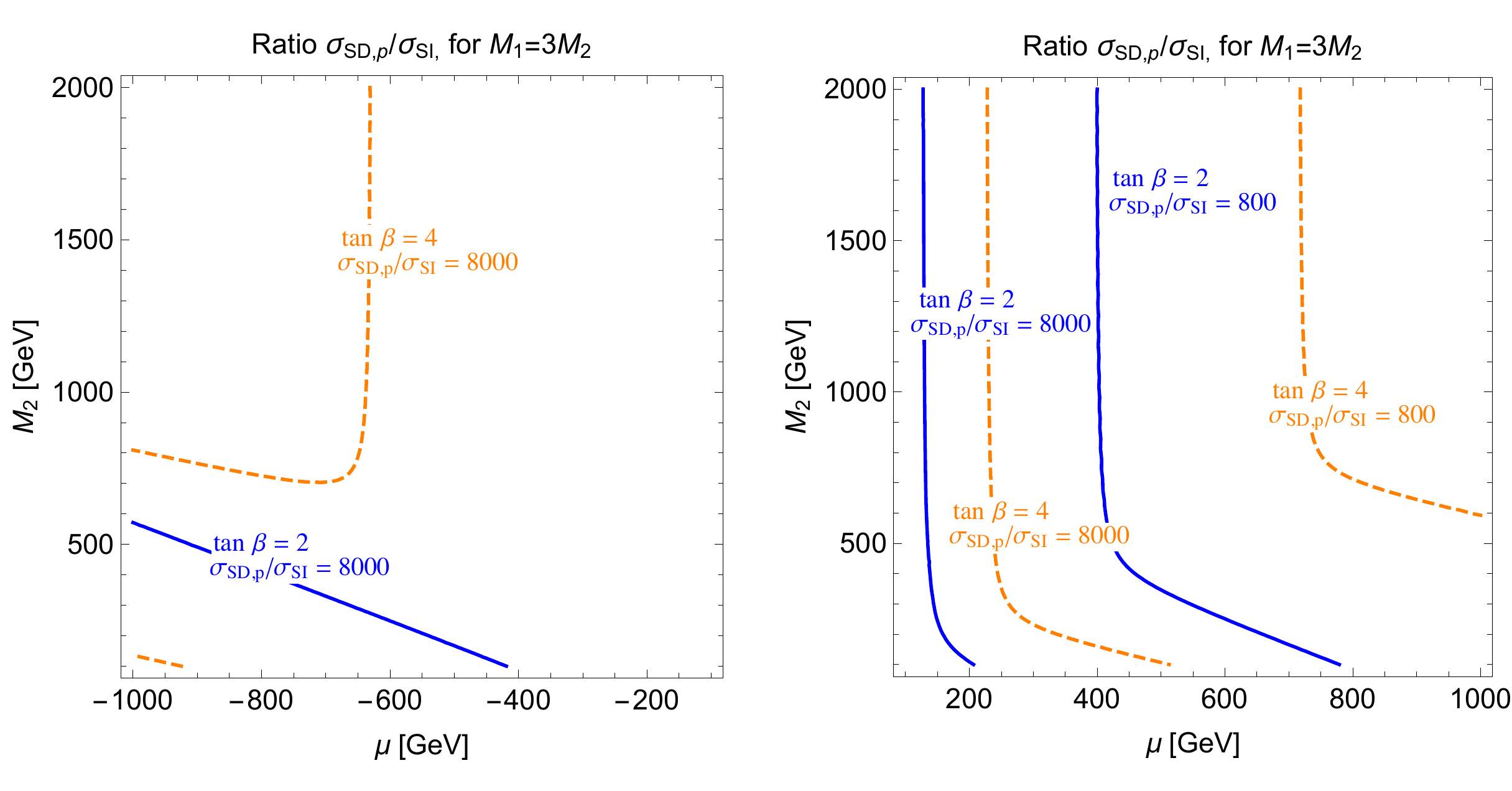}
\end{center}
\caption{Contours of the ratio of the spin-dependent scattering cross section on protons, $\sigma_{{\rm SD},p}$, to the spin-independent scattering cross section $\sigma_{\rm SI}$ in the $(\mu, M_2)$ plane with the choice $M_1 = 3 M_2$ (as in AMSB). At left, $\mu < 0$; at right, $\mu > 0$. Spin-independent scattering rates are larger when $\mu$ is positive. The dashed contours are for $\tan \beta = 4$ and the solid contours for $\tan \beta = 2$. We see that typically the spin-dependent cross section is several thousand times the spin-independent one. 
}
\label{fig:directdetection}
\end{figure}%

\begin{figure}[!h]\begin{center}
\includegraphics[width=0.65 \textwidth]{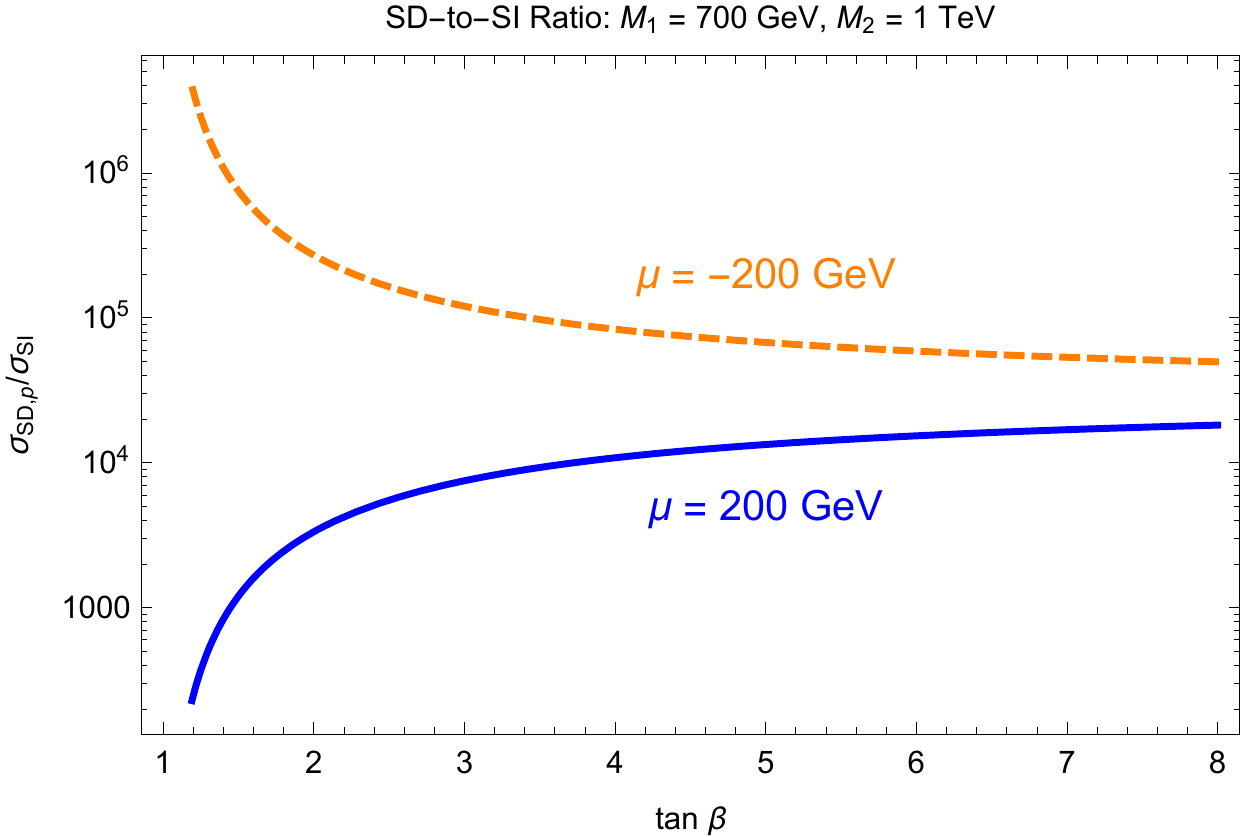}
\end{center}
\caption{The $\tan \beta$ dependence of the cross-section ratio $\sigma_{{\rm SD},p}/\sigma_{SI}$ for a point in parameter space ($M_1 = 700$ GeV, $M_2 = 1000$ GeV, $|\mu| = 200$ GeV). At left, $\mu < 0$; at right, $\mu > 0$. When $\mu > 0$, the ratio increases by a factor of 3.2 as $\tan \beta$ increases from 2 to 4.}
\label{fig:directdetectionSDvsSI}
\end{figure}%

Although the scattering cross sections $\sigma_{\rm SI, SD}$ are {\em individually} $\tan \beta$-dependent, they are not directly measurable. The local density of dark matter is known only to $\sim 30\%$ accuracy (see e.g.~\cite{Bovy:2012tw}); even if this improves in the future, we will not necessarily know that the dark matter particle detected in an experiment makes up {\em all} of the local dark matter density. Hence, the ratio $\sigma_{\rm SD}/\sigma_{\rm SI}$ is a more useful analyzer of $\tan \beta$, since astrophysical uncertainties cancel in this ratio.

The relationship between the expected size of spin-dependent and spin-independent signals for neutralino dark matter has been discussed extensively in \cite{Cohen:2010gj}. Currently some of the strongest constraints on spin-independent scattering come from LUX \cite{Akerib:2016vxi} and PandaX-II \cite{Tan:2016zwf}; for spin-dependent scattering, from IceCube (in the case of protons) \cite{Aartsen:2016exj} and PandaX-II (in the case of neutrons) \cite{Fu:2016ega}. For a WIMP mass of 200 GeV, the current bounds are roughly $\sigma_{\rm SI} \lesssim 3 \times 10^{-46}~{\rm cm}^2$ and $\sigma_{\rm SD} \lesssim 10^{-40}~{\rm cm}^2$. From \eqref{eq:directrates} we see that these probe roughly similar values of $c_{h \chi \chi}$ and $c_{Z \chi \chi}$, but our theoretical expectation is that $c_{Z \chi \chi}$ is typically smaller (at least for $\mu > 0$). We have illustrated the expected relative size $\sigma_{{\rm SD},p}/\sigma_{\rm SI}$ in Fig.~\ref{fig:directdetection}, and the $\tan \beta$ dependence of this ratio in Fig.~\ref{fig:directdetectionSDvsSI} for a particular choice of masses. In the latter plot we see that when $\mu > 0$, typically the spin-dependent cross section is larger by a factor of $\sim 10^3$ at low $\tan \beta$ and $\sim 10^4$ at large $\tan \beta$. When $\mu < 0$, the spin-dependent scattering rate is larger by $\sim 10^5$, and increasingly large relative to the spin-independent rate as $\tan \beta \to 1$. 

The spin-dependent to spin-independent cross section ratio may be a powerful probe of $\tan \beta$, but this requires some optimism. We can hope for a spin-independent signal in a near future experiment, at the $\sim 10^{-46}~{\rm cm}^2$ level. (This may occur at a point in parameter space for which $\sigma_{\rm SI}$ itself is larger, but the neutralino constitutes only a fraction of the dark matter, so that the effective $\sigma_{\rm SI}$ inferred from the experiment is smaller.) Then spin-dependent tests must probe small cross sections of order $10^{-43}$ to $10^{-42}~{\rm cm}^2$ in order to measure (or at least put an informative upper bound on) the ratio $\sigma_{\rm SD}/\sigma_{\rm SI}$. For instance, at the point in parameter space shown in the right panel of Fig.~\ref{fig:directdetectionSDvsSI}, which at $\tan \beta = 4$ has $\sigma_{\rm SI} \approx 2.5 \times 10^{-46}~{\rm cm}^2$, a measurement of a ratio $\sigma_{{\rm SD},p}/\sigma_{\rm SI} = (10 \pm 1) \times 10^{3}$ would determine $\tan \beta = 3.7 \pm 0.3$. The Snowmass working group report on direct detection suggests that bounds of $\sigma_{\rm SD} \lesssim {\rm few} \times 10^{-42}~{\rm cm}^2$ may be achieved by LZ and PICO250 \cite{Cushman:2013zza}, but does not forecast any improvements beyond this. We would argue that a positive signal consistent with spin-independent scattering in future direct detection experiments would strongly motivate an intense effort to achieve another order of magnitude or two improvement in spin-dependent scattering in order to measure the ratio $|c_{Z\chi\chi}/c_{h\chi\chi}|$ and hence, in the MSSM context, $\tan \beta$.

Kinematic measurements at a collider can tell us $M_1$, $M_2$, and $|\mu|$, but are less sensitive to the sign of $\mu$. However, notice from Fig.~\ref{fig:directdetectionSDvsSI} that the range of ratios $\sigma_{{\rm SD},p}/\sigma_{\rm SI}$ for positive and negative $\mu$ do not overlap. This means that a measurement of the spin-dependent to spin-independent scattering rate can simultaneously be used to measure $\tan \beta$ and $\sign(\mu)$. However, $\sigma_{\rm SI}$ is so small when $\mu < 0$ that a successful measurement of this ratio may be much more challenging.

\section{A 100 TeV collider case study with \texorpdfstring{$\bm{M}_2 > |\bm{\mu}| > \bm{M}_1$}{$M_2 > |\mu| > M_1$}}
\label{sec:colliderstudy}

We will present collider studies for a proton--proton collider operating at $\sqrt{s} = 100$ TeV measuring $m_0$ and $\tan \beta$ for a benchmark model with the following parameters for the gaugino and higgsino sector: 
\begin{align}
M_3 = 2~{\rm TeV}, \quad M_2 = 800~{\rm GeV}, \quad M_1 = 200~{\rm GeV}, \quad {\rm and} \quad \mu = 400~{\rm GeV}. 
\end{align}
 In this case, the bino is the LSP. The second and third heavier neutralinos, $\widetilde{\chi}_2^0$ and $\widetilde{\chi}_3^0$, are the higgsinos, while the heaviest neutralino is the wino. Among the two neutral higgsinos, $\widetilde{\chi}_3^0$ decays dominantly to $Z$ bosons and LSPs while $\widetilde{\chi}_2^0$ goes dominantly to higgses and LSPs. 

There are two sources of background: Standard Model backgrounds, which can mostly be removed by hard cuts on missing $p_T$ and $H_T$; and SUSY backgrounds, i.e.~confusion among different decay modes. 
In simulating signal events we use Pythia \cite{Sjostrand:2014zea} supplied with a decay table computed by SUSY-HIT \cite{Djouadi:2006bz} and modified to include gluino decays as computed in \cite{Gambino:2005eh} (which includes the resummation of the radiative corrections). In studies of Standard Model backgrounds, we have also used MadGraph \cite{Alwall:2014hca}, MadSpin \cite{Artoisenet:2012st}, and MLM matching \cite{Alwall:2007fs}. We use leading order simulations (but including matching of one or two extra jets where appropriate) and rescale the cross sections reported by MadGraph and Pythia to match the most accurate NLO or NNLO results in \cite{Borschensky:2014cia, Beenakker:1999xh, Mangano:2016jyj, Contino:2016spe} for a given process. Jets are clustered using FastJet \cite{Cacciari:2005hq,Cacciari:2011ma} and the anti-$k_t$ algorithm \cite{Cacciari:2008gp}. 

Studies of future hadron colliders are still at an early stage, so basic questions about what rapidity cuts, trigger thresholds, identification efficiencies, or energy resolutions to consider are still open. Hence we forego detector simulation and make some simple pragmatic choices. Early studies of 100 TeV colliders have made a case for having a significantly extended pseudorapidity coverage relative to the LHC \cite{Mangano:2016jyj, Contino:2016spe}. This is readily understood: in a process with partonic center-of-mass energy $E$, the largest accessible rapidities for particles of mass $m \lesssim E$ are $\sim \log(E/m)$. Rapidity distributions will be fairly flat up to this point. Increasing $E$ by an order of magnitude raises the maximum accessible rapidities by roughly 2. This is borne out by a number of plots of Standard Model processes in \cite{Mangano:2016jyj}. Hence we assume that the design of a detector for future hadron colliders will have an extended pseudorapidity coverage compared to the LHC, in order to not sacrifice efficiency and hermeticity for Standard Model measurements. To that end, we assume efficient object identification in the ranges
\begin{align}
\left|\eta_{\rm jet}\right| & \leq 5, \nonumber \\
\left|\eta_{\rm lepton}\right| & \leq 3.5.
\end{align}
We also require the leptons to have $p_T > 10$ GeV. In addition, the total sum of $p_T$ of all the charged tracks within a cone of radius 0.3 around the lepton have to be smaller than 15\% of the lepton's $p_T$.

\subsection{Measuring \texorpdfstring{$\bm{m}_0$}{$m_0$}}
\label{ssec:m0}
When we vary the scalar mass $m_0$ from 30 TeV to 1000 TeV, the two
body branching fraction ${\rm Br} (\widetilde g \to
\widetilde{\chi}_3^0 g)$ increases from 1\% to 2.4\%, due to the
logarithmic sensitivity discussed in \S\ref{sec:scalarmassmeasure}.
Below we will present a simple set of cuts that could give us a sample
with a considerable fraction of events with at least one two-body
decaying gluino, which allows us to measure $m_0$. 

\begin{figure}[h]
  \centering
  \includegraphics[width=0.65\textwidth]{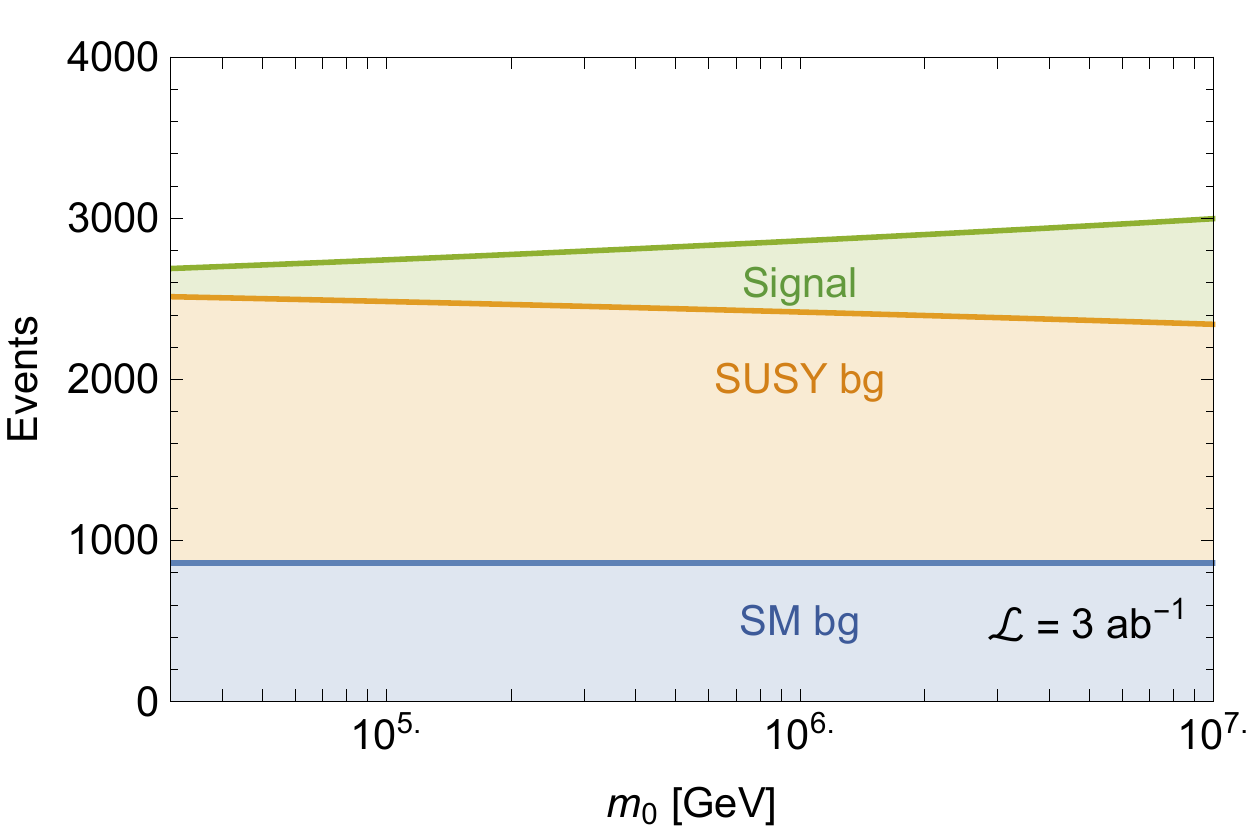}
  \caption{
    Stacked number of events passing our cuts at 3 ab$^{-1}$ for the signal and
  background as a function of $m_0$.
  We define the signal as events with at least one two-body decaying gluino. Events with two three-body decaying gluinos are SUSY backgrounds. The SM
  background mainly consists of $ZZ$ + jets and $t\bar{t}+Z$ production.
  We use NLO production 
  cross sections for the signal~\cite{Borschensky:2014cia} and background~\cite{Mangano:2016jyj}.  }
  \label{fig:eventsm0}
\end{figure}

\begin{figure}[h]
\centering
\includegraphics[width=1.0\textwidth]{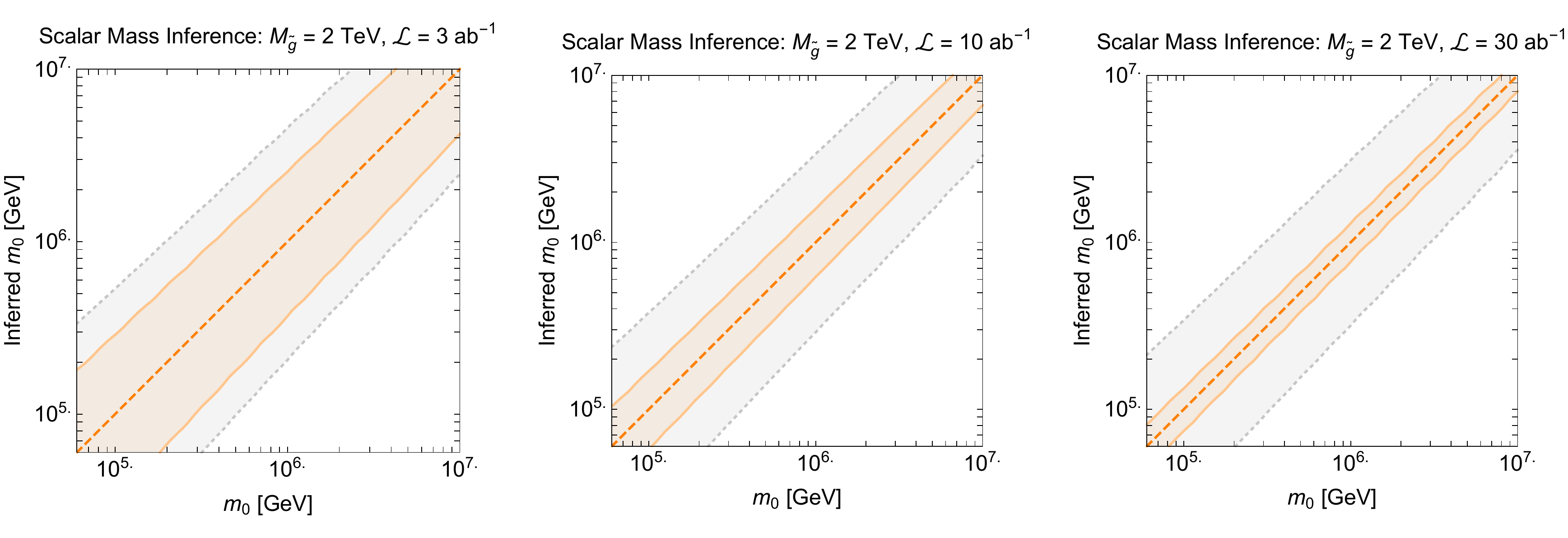}
\caption{Inference of the scalar mass scale $m_0$ from the measurement of the rate of 2-body decays $\tg \to g \tH^0$. The parameters are $M_3 = 2$ TeV, $\mu = 400$ GeV, $M_1 = 200$ GeV, and $M_2 = 800$ GeV. The orange band represents $1\sigma$ statistical uncertainty with 3 ab$^{-1}$ of data (left), 10 ab$^{-1}$ of data (middle) and 30 ab$^{-1}$ of data (right), while the grey band corresponds to a 3\% systematic uncertainty on cut efficiencies times cross section times luminosity.}
\label{fig:M0EstimationPoint1}
\end{figure}%

The set of cuts we adopt are: 
\begin{align}
H_T &> 2~{\rm TeV},  \quad\quad p_T^{\rm missing} > 1~{\rm TeV}, \quad\quad p_T(j_1) > 1~{\rm TeV},\\
N_{\rm jet} &< 5, ~{\rm one~leptonic~}Z~(80~{\rm GeV} < m_{\ell\ell} < 100~{\rm GeV}), \\
m_{j_1Z} &> m_{\rm all~other~jets}, ~M_{T2}^{\ell\ell} > 80~{\rm GeV}. 
\end{align}
The jets are clustered with $R = 0.6$ and required to satisfy $\left|\eta \right|< 3.5$ and $p_T > 100$ GeV. $H_T$ is the scalar sum of the jet $p_T$. $j_1$ denotes the hardest (i.e.~highest $p_T$) jet. Since $\widetilde{\chi}_3^0$ from the gluino two-body decays subsequently decays to $Z$ plus LSP, we require that there are at least two leptons in the event with one opposite-sign same-flavor pair reconstructing a $Z$ boson. Events with at least one two-body decaying gluino tend to have fewer jets and a larger invariant mass of the leading jet and the $Z$ boson compared to events in which both gluinos decay through three-body processes. These features are reflected by the cuts on the number of jets and on the ratio between the invariant mass of the leading jet and $Z$ and that of all the other jets.
Standard Model backgrounds in which missing energy arises dominantly from neutrinos in $W^+ W^-$ or $t \overline{t}$ decays can be rejected by the subsystem $M_{T2}$ variable built out of the two leptons and missing $p_T$ \cite{Burns:2008va}, which we denote $M_{T2}^{\ell\ell}$. This ``dileptonic $M_{T2}$'' variable generalizes the original inclusive $M_{T2}$ \cite{Lester:1999tx} and has been discussed as a useful $t \overline{t}$ rejector in SUSY searches in \cite{Kats:2011qh, Kilic:2012kw}. We calculate $M_{T2}^{\ell\ell}$ using the code distributed with \cite{Lester:2014yga}. 

With these cuts, we found that for events with at least one $\widetilde g \to \widetilde{\chi}_3^0 g$, the efficiency of the cuts (the fraction of events that passes cuts) is $3.6 \times 10^{-4}$. For events with one $\widetilde g \to \widetilde{\chi}_2^0 g$ (and one gluino three-body decay), the efficiency is $6.5 \times 10^{-5}$. These two classes of events are counted as signals. The SUSY background comes from events with two gluino three-body decays and has an efficiency $7.8 \times 10^{-5}$. Given these efficiencies, for $3$ ab$^{-1}$ luminosity, there are $\sim 1600$ SUSY background events as well as about 860 Standard Model events. The dominant Standard Model background is $Z(\to \ell^+ \ell^-) + Z(\to \nu \bar{\nu}) + {\rm jets}$, which contributes about 560 events, while $t{\bar t} + Z(\to \ell^+ \ell^-)$ contributes about 300 events. The $t{\bar t} + {\rm jets}$ background is negligible in comparison, though prior to the $M^{\ell \ell}_{T2}$ cut it was dominant. The number of SUSY signal events varies from $175$ at $\tan \beta = 4$ to 450 at $\tan\beta = 2$. The number of events passing cuts as a function of $m_0$ is presented in Fig.~\ref{fig:eventsm0}.

The estimated performance of a simple cut-and-count analysis is presented in Figure \ref {fig:M0EstimationPoint1}. The orange band shows that statistical uncertainty alone can be quite small. The gray band represents an additional 3\% systematic uncertainty in the event rate. This corresponds to about a factor of $\sim 5$ uncertainty in the scalar mass scale, bracketing $\Smiley_L$ to the range $8.1-180$ TeV and $\Smiley_H$ to $204-4620$ TeV. It seems likely a multivariate analysis will outperform our simple cuts, reducing our sensitivity to systematic uncertainties.

\subsection{Measuring tan \texorpdfstring{$\bm{\beta}$}{$\beta$}}
\label{ssec:tb}

For the benchmark point, we scan over $\tan\beta$ values in the range (2.0,4.0).
The analysis to extract $\tan\beta$ relies on higgsino pair production
and subsequent decay to bino in addition to a $Z$ or a higgs, which is $\tan \beta$ dependent for reasons discussed in \S\ref{subsec:blindspot}.

The basic signal is a pair of $Z$-bosons in addition to $p_T^{\rm missing} $.
The event selection is as follows,
\begin{enumerate}
  \item Two pairs of opposite sign same flavor leptons, with
    $|m_{\ell \ell} - m_Z | < 10 \GeV$.
  \item $p_T^{\rm missing} > 150 \GeV$.
  \item Scalar sum of $p_T$ of all visible particles $< 600 \GeV$.
\end{enumerate}

\begin{figure}[t]
  \centering
  \includegraphics[width=0.65\textwidth]{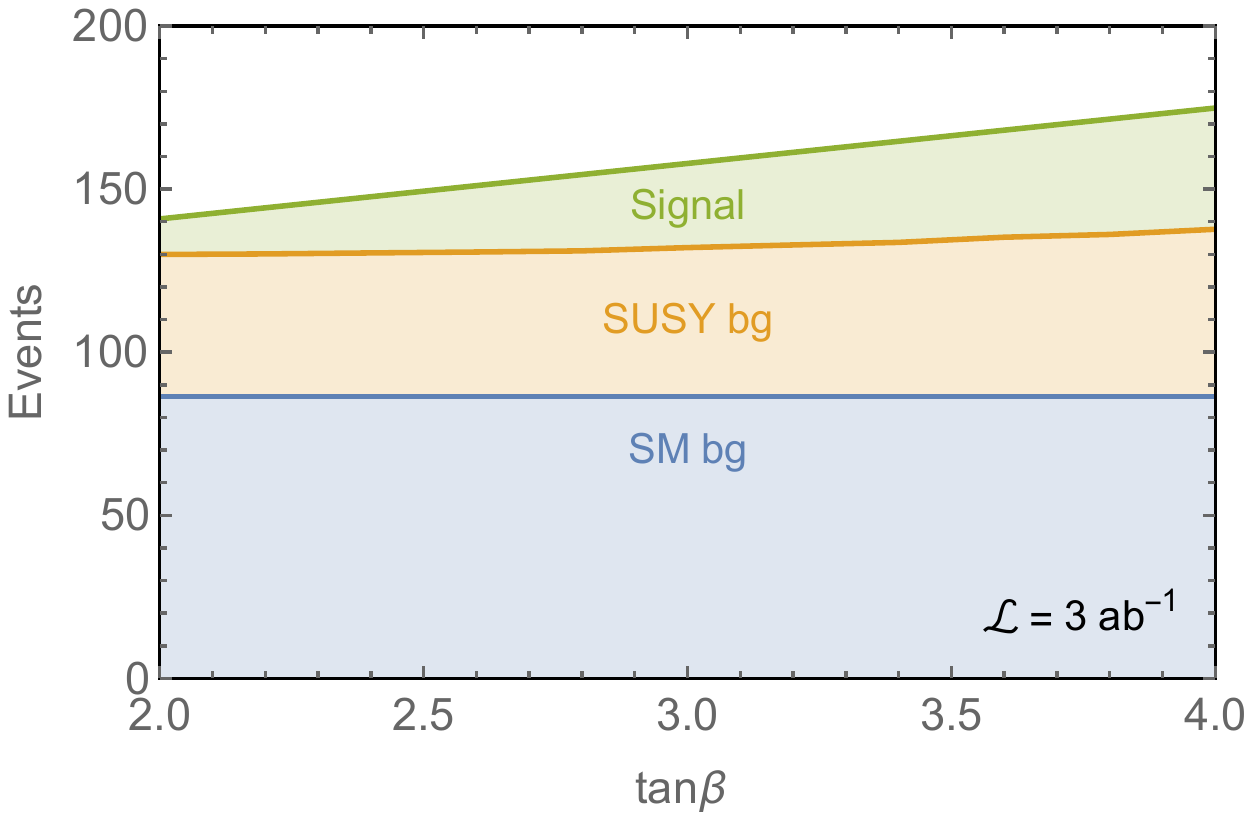}
  \caption{
    Stacked number of events passing our cuts at 3 ab$^{-1}$ for the signal and
  background for different values of $\tan\beta$.
  We define the signal as $(\tilde{\chi}_2^0 \to Z
 \tilde{\chi}_1^0)(\tilde{\chi}_3^0 \to Z \tilde{\chi}_1^0)$. All other neutralino cascades
  are considered to be a part of the SUSY background. The SM
  background mainly consists of $ZZZ$ production.
  We use NLO production 
  cross sections for the signal (Prospino 2~\cite{Beenakker:1999xh})
  and background~\cite{Mangano:2016jyj}.
  }
  \label{fig:point3eff}
\end{figure}

\begin{figure}[h]
  \centering
  \includegraphics[width=0.3\textwidth]{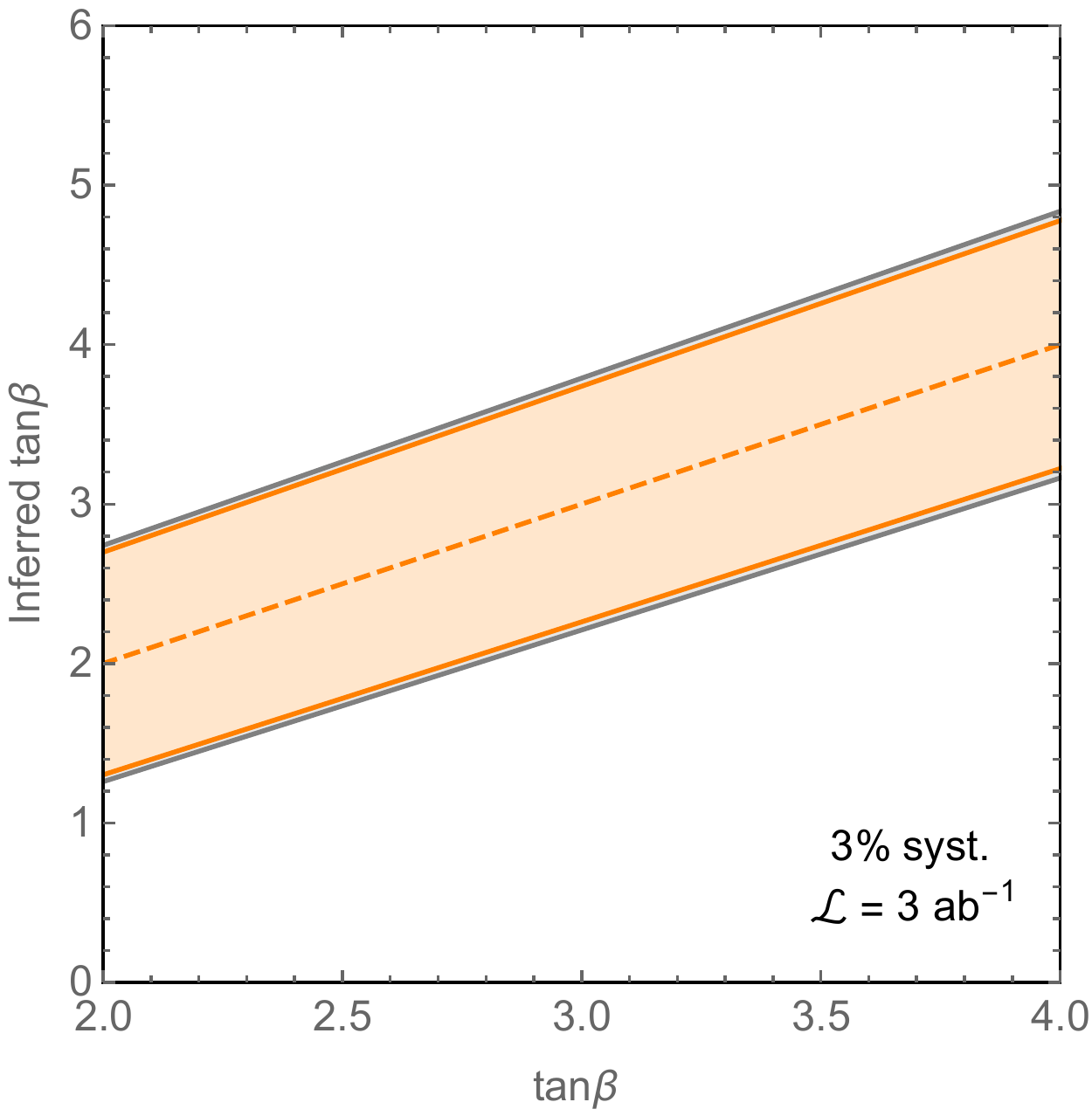}
  \quad
  \includegraphics[width=0.3\textwidth]{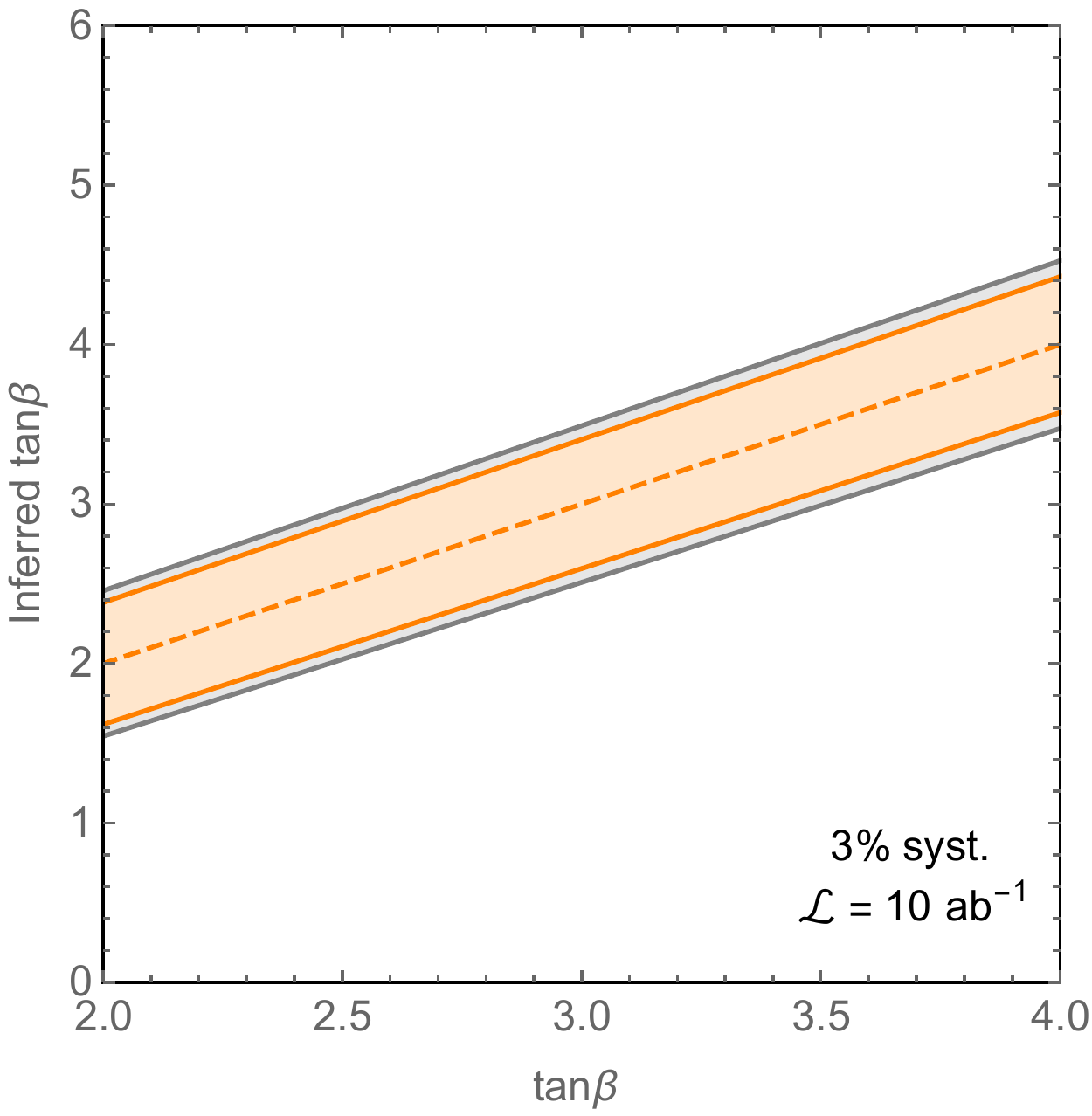}
  \quad
  \includegraphics[width=0.3\textwidth]{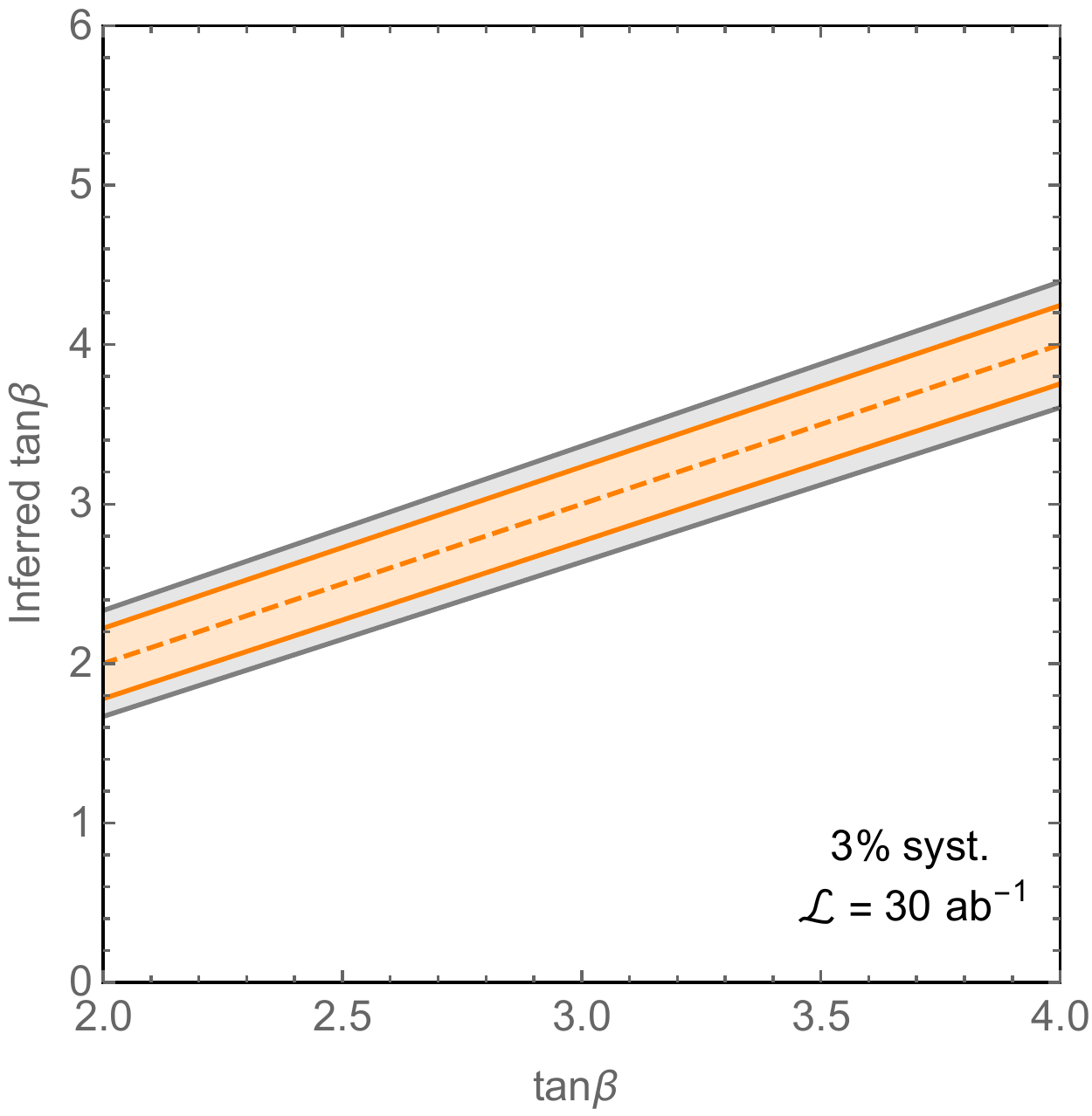}
  \caption{Inference of $\tan \beta$ from the measurement of
    $\widetilde{H}^0 \rightarrow Z \widetilde{B} $ decays. The
    parameters are $M_3 = 2 ~\mathrm{TeV}$, $M_1 = 200
    ~\mathrm{GeV}$, $M_2 = 800 ~\mathrm{GeV}$, $\mu =
    400~\mathrm{GeV}$. The orange shaded band represents $1\sigma$
    statistical uncertainty and the gray region additionally includes
    a $3\%$ systematic uncertainty. We show results for
    luminosity values $\mathcal{L}=3,10,30~\mathrm{ab}^{-1}$.
  }
  \label{fig:infer3-new}
\end{figure}

For this analysis, jets are clustered using the anti-$k_t$ algorithm,
using a jet
radius of $R=0.4$, and are required to have $|\eta| < 5$. The third cut is employed to reduce SUSY background from neutral and
charged wino, as well as gluino, production.

The dominant SM background arises from $ZZZ$
production. There is a potential background from $hZ$ production, but
it has negligible efficiency for our analysis. We use the NLO SM cross
sections reported in~\cite{Mangano:2016jyj}. 
We used Prospino 2 to calculate the NLO cross sections for
electroweakino pair
production~\cite{Beenakker:1999xh}. The total number of
events passing for each case is shown in Figure~\ref{fig:point3eff}.

The efficiency for the SM background
$p p \to Z Z Z \to (\ell^+ \ell^-)^2 \nu\nu$ is $0.046$, resulting in
$87$ SM background events passing all cuts. The efficiency for
the signal $ p p\to\tchi_2^0 \tchi_3^0 \to (Z\to\ell^+\ell^-)^2  \tchi_1^0
\tchi_1^0$ is $0.15$, which translates into
$11$--$37$ events over the $\tan\beta$ range. There are a number of
different channels contributing to 
the SUSY background, together yielding $41$--$51$ events. For
these numbers above we have assumed a luminosity
$\mathcal{L}=3\ \mathrm{ab}^{-1}$.
The inferred
value of $\tan\beta$ as a function of the Monte Carlo truth is shown
in Figure~\ref{fig:infer3-new}.

\subsection{The origin of the higgs mass}
\begin{figure}[!h]
  \centering
  \includegraphics[width=0.65\textwidth]{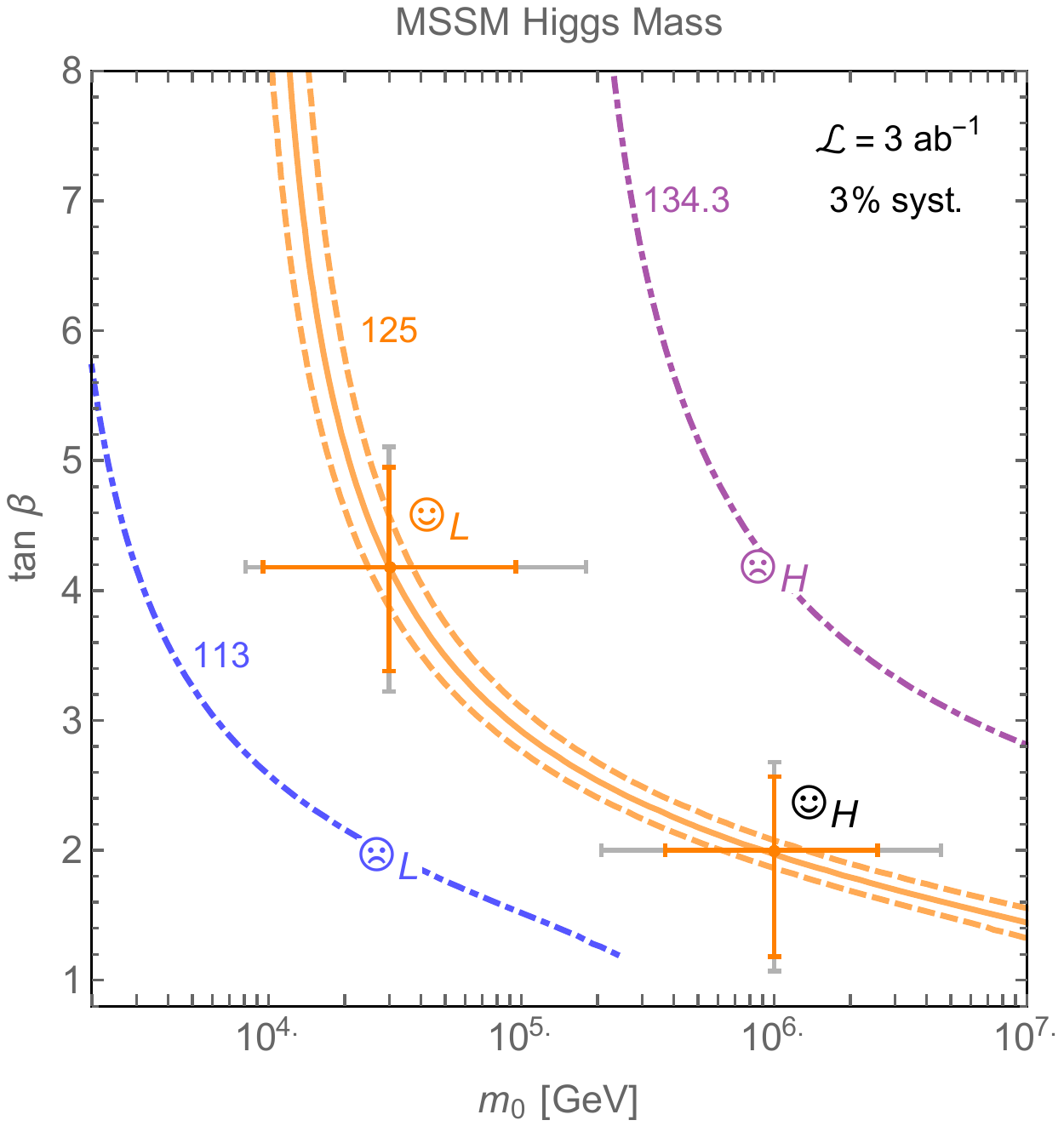}
  \caption{Expected accuracy in measurement of $m_0$ and $\tan\beta$
    for two parameter points consistent with $m_h=125\GeV$ within the
    MSSM. The error bars correspond to $1\sigma$ statistical
    uncertainty (orange) and an additional $3\%$ systematic
    uncertainty (gray). We chose the luminosity to be
    $\mathcal{L}=3~\mathrm{ab}^{-1}$.
  }
  \label{fig:m0tbinf}
\end{figure}

We use our results from sections~\ref{ssec:m0} and \ref{ssec:tb}
and overlay them with the higgs mass contours in the MSSM in
figure~\ref{fig:m0tbinf}. We see that with our simple analysis it is
indeed possible to distinguish the four benchmark points
($\Smiley_{L,H},\Sadey_{L,H}$) at $\sim 2\sigma$ level. It
is interesting that even within the MSSM we can distinguish between
the higher and lower scalar mass scales, which can give us additional
information about SUSY breaking.

\section{Conclusions and Outlook}
\label{sec:conclusions}

The discovery of the higgs boson and the measurement of its mass at
the LHC provide qualitatively new information about the Standard Model. The
mass $m_h= 125$ GeV is intriguingly close to the range below $m_Z$ predicted
in the minimal supersymmetric model, with the difference arising from
quantum corrections to the higgs quartic. Future colliders will have
the potential to test if the origin of the higgs mass is indeed from
such quantum corrections. This can be a powerful test of the MSSM, and if the stop quantum corrections are
not by themselves responsible for the higgs mass, then it would point
towards an extended model
like the NMSSM (which involves a new singlet superfield) or new U(1)
gauge symmetries (which can provide new D-term contributions). These
theories might also come with additional correlated signatures from
these additional particles. Even in the split SUSY limit, these
modified theories will contain new light fermions.

In this work we have constructed a list of
observables which have the potential to measure the parameters $m_0$
and $\tan\beta$. We focused on a theoretically motivated but
experimentally challenging part of parameter space, where only the
gauginos and higgsinos are light enough to be
produced and the scalar superpartners
are in the mass range $30$--$1000$ TeV, too heavy to be directly
accessible to a 100 TeV
collider. The $\tan\beta$ values were correspondingly chosen to be between 2 and 4. 

Further,
we picked a specific benchmark spectrum and performed a
detailed collider study. 
Loop-mediated two-body gluino decays can be used to
measure the scalar mass scale within a factor of 5 (at 3 ab$^{-1}$
assuming 3\% systematic uncertainty). Pair production of higgsino-like
NLSP states with subsequent decay to LSPs and a pair of $Z$ bosons can
help measure $\tan\beta$ within $\pm0.8$ with the same assumptions as
above. We showed that a combination of these observables indeed have the potential to test
the MSSM origin of the higgs mass. 

We intend our analysis to be a first proof of principle, and a number of
improvements can be easily imagined. A wider swath of SUSY parameter
space could be explored, including heavier gluino masses (which have
lower production rates, but are kinematically more distinct from the
Standard Model backgrounds). While we have focused on very clean
leptonic channels (at the cost of reducing the signal strength
due to small branching ratios), it is plausible that using top- and
$W,Z$-tagging will increase sensitivity to the signals. In addition,
our simple cut and count based analyses could certainly be improved by
multivariate analysis and even more sophisticated tools, e.g., from
deep learning. Another interesting future direction is to assess the
impact of different collider energies and luminosities on how well we
can test the origin of the higgs mass within the MSSM. To fairly
compare the reach, we should standardize an analysis procedure that
works across energies, rather than choosing cuts by hand; multivariate
analyses trained in the same way on different input data may lend
themselves well to this. The interplay with other future experiments
is another important avenue to understand better. We have sketched how
dark matter direct detection experiments could provide one such
important source of complementary information. In order to make use of
direct detection, it is important to further improve the prospects for
measuring spin-dependent scattering at low cross sections. 

Our study has served as one example of how investigating a particular physical mechanism, rather than pure discovery reach for particles, can lead to specific targets for colliders. Further studies aimed at a variety of mechanisms will help to inform the design of future collider experiments.

\section*{Acknowledgments}

We thank Matthew Low, Olivier Mattelaer, Michelangelo Mangano, Matthew McCullough, Brian Shuve, and Neal Weiner for useful discussions or correspondence. MR thanks the Institute for Advanced Study for funding and hospitality while this work was completed. PA is supported by NSF grant PHY-1216270. JF is supported by the DOE grant DE-SC-0010010. MR is supported in part by the NSF Grant PHY-1415548 and the NASA grant NNX16AI12G. WX is supported by the U.S. Department of Energy (DOE) under cooperative research agreement DE-SC-00012567. Some computations in this paper were run on the Odyssey cluster supported by the FAS Division of Science, Research Computing Group at Harvard University.

\bibliography{ref}
\bibliographystyle{utphys}
\end{document}